\documentclass[twocolumn,tighten]{aastex63}

\usepackage{times,natbib,graphicx,amsmath,multirow,xspace}
\usepackage{booktabs}
\usepackage{placeins}
\usepackage{float}

\usepackage{xcolor}
\usepackage{lineno}
\let\tablenum\relax
\usepackage{siunitx}
\usepackage{hyperref}
\usepackage{tabularx}
\usepackage{longtable}

\newcommand{\nustar}{NuSTAR\xspace}

\newcommand{\fluxcgs}{ergs~cm$^{-2}$~s$^{-1}$\xspace}
\newcommand{\lumcgs}{ergs~s$^{-1}$\xspace}
\newcommand{\rin}{$R_{\rm in}$\xspace}
\newcommand{\rg}{$R_{g}$\xspace}
\newcommand{\risco}{$R_{\mathrm{ISCO}}$\xspace}

\newcommand{\relxillns}{{\sc relxillNS}\xspace}

\newcommand{\source}{\mbox{GX 340+0}\xspace}


\shorttitle{GX 340+0: In and Out of Focus}
\shortauthors{Li et al.}

\begin{document}

\title{GX 340+0: In and Out of Focus}

\author{S.~Li}
\affiliation{Department of Physics \& Astronomy, Wayne State University, 666 West Hancock Street, Detroit, MI 48201, USA}

\author[0000-0002-8961-939X]{R.~M.~Ludlam}
\affiliation{Department of Physics \& Astronomy, Wayne State University, 666 West Hancock Street, Detroit, MI 48201, USA}

\author[0000-0002-5341-6929]{D.~J.~K.~Buisson}
\affiliation{Independent Researcher}

\author[0000-0003-0440-7978]{M.~Sudha}
\affiliation{Department of Physics \& Astronomy, Wayne State University, 666 West Hancock Street, Detroit, MI 48201, USA}

\author[0000-0001-6304-1035]{S.~Rossland}
\affiliation{Independent Researcher}

\author[0000-0003-4216-7936]{G.~Mastroserio}
\affiliation{Dipartimento di Fisica, Universit\'a Degli Studi di Milano, Via Celoria, 16, Milano, 20133, Italy}

\author[0000-0002-4024-6967]{M.~C.~Brumback}
\affiliation{Department of Physics, Middlebury College, Middlebury, VT 05753, USA}

\author[0000-0003-3828-2448]{J.~A.~Garc\'{i}a}
\affiliation{X-Ray Astrophysics Laboratory, NASA Goddard Space Flight Center, Greenbelt, MD, USA}
\affiliation{Cahill Center for Astronomy and Astrophysics, California Institute of Technology, Pasadena, CA 91125, USA}

\author[0000-0002-1984-2932]{B.~W.~Grefenstette}
\affiliation{Cahill Center for Astronomy and Astrophysics, California Institute of Technology, Pasadena, CA 91125, USA}

\author[0000-0001-8916-4156]{F.~La Monaca}
\affiliation{INAF Istituto di Astrofisica e Planetologia Spaziali, Via del Fosso del Cavaliere 100, 00133 Roma, Italy}
\affiliation{Dipartimento di Fisica, Universit\`{a} degli Studi di Roma ``Tor Vergata'', Via della Ricerca Scientifica 1, I-00133 Roma, Italy}

\author[0000-0003-4841-8302]{E.~A.~Saavedra}
\affiliation{Instituto de Astrof\'isica de Canarias (IAC), V\'ia Láctea, La Laguna, E-38205, Santa Cruz de Tenerife, Spain}
\affiliation{Departamento de Astrof\'isica, Universidad de La Laguna, E-38206, Santa Cruz de Tenerife, Spain}

\author[0000-0003-0331-3259]{A.~Di Marco}
\affiliation{INAF Istituto di Astrofisica e Planetologia Spaziali, Via del Fosso del Cavaliere 100, 00133 Roma, Italy}

\begin{abstract}
The Nuclear Spectroscopic Telescope Array (\nustar) enables detailed high-energy X-ray observations from 3--79 keV, but its performance can be constrained by telemetry saturation when observing bright sources, leading to reduced effective exposure times. 
In this study, we investigate the use of serendipitous stray light (SL) observations to infer properties of an X-ray bright source in comparison to focused data. 
Our case study is performed on the neutron star (NS) low-mass X-ray binary (LMXB) GX 340+0, a prominent Z source, where we execute a spectral analysis comparing 25 SL and 7 focused \nustar observations. 
Our findings demonstrate that SL observations can significantly enhance long-term temporal coverage; detecting variations in the thermal components of the system across the baseline of the mission, which could not be inferred from focused observations alone.
\end{abstract}


\section{Introduction} \label{sec:intro}

The Nuclear Spectroscopic Telescope Array (\nustar: \citealt{Harrison_2013}) is a NASA telescope designed to observe high energy X-rays with two focal plane modules (FPMA and FPMB) of identical construct. Unlike many other X-ray telescopes that focus low energy X-rays below 10 keV, the multi-layer mirror coatings composed of high-density and low-density materials (Pt/C and W/Si) enhance reflectivity of X-ray up to 79 keV effectively allowing \nustar to operate in the energy range of 3--79 keV. Targeted observations from \nustar have been filling critical gap in X-ray astronomy, specifically in the area of high spatial resolution observations above 10 keV. 
However, sometimes when observing bright sources, the incoming photons may result in extremely high count rates, causing the observations to have a shorter exposure time due to dead time and high telemetry loads \citep{Grefenstette_2021}. However, \nustar has an open geometry between the optics and detectors which allows light from off-axis sources (about 1--4 degrees) to pass through without being focused by the optics \citep{straylight}. Light falling on the detector but not focused by the optics are referred to as ``stray light'' (SL). Because SL observations are usually not intended but rather unwanted or extraneous light for targeted observations of other sources \citep{Harrison_2013}, they do not take any extra telescope time and can potentially have more data in comparison to targeted observations of a single source. These observations have been collected and compiled into a catalog of stray light observations known as `StrayCats' \citep{Grefenstette_2021, ludlam22b}

Since SL observations are not focused through the optics, they have lower count rate and signal to noise ratio (SNR) in comparison, which makes the spectral analysis more involved than focused observations. Consequently, SL observations are usually better served when studying bright sources. In fact, some SL observations have been intentionally performed for bright sources to reduce the telemetry load. Additionally, for spectrally hard X-ray sources observed as intentional SL, the photons collected are not limited by the sensitivity of the focusing optics and the spectra can be extended to energies beyond 79 keV as long as the background does not dominate \citep{Mastroserio_2022}.

Low-mass X-ray binaries (LMXBs) are often among the brightest X-ray sources in the sky. They consist of a star of $\lesssim1\ M_{\odot}$ and a compact object: a neutron star (NS) or black hole (BH). Matter is stripped from the stellar companion via Roche-lobe overflow to form an accretion disk around the compact object. For NS LMXBs, based on the X-ray spectral and timing properties, they are divided into two categories: `Z' sources and `atoll' sources \citep{Van_1989}. These two categories are named after the shape of their color-color or hardness intensity diagrams (HID). Atoll sources have an ``island-like'' shape on their HID and they are divided into two states: the hard `island' state and soft `banana' state. Z sources have a ``Z'' shape and are divided into three different branches: horizontal branch (HB), normal branch (NB), and flaring branch (FB). Additionally, the transition from HB to NB is called the hard apex (HA) and the transition from NB to FB is called the soft apex (SA).  

The spectra of NS LMXBs are generally soft and thermally dominated, thus are commonly modeled using a combination of multiple thermal components.
\cite{Mitsuda_1984} suggested that every observed spectrum can be expressed by a soft and a hard thermal component. The hard component is a 2 keV blackbody spectrum describing the NS surface and the soft component is a multi-color blackbody spectra describing the accretion disk. Later on in the 1980s, two different approaches were proposed to account for Comptonization in LMXBs: the `Eastern' model \citep{eastern} and the `Western' model \citep{western}. 
The Eastern model described the spectrum with multi-color disk component and Comptonization arising from a single-temperature blackbody attributed to the boundary layer (BL), where the accretion disk material is decelerated and heated before settling onto the NS surface \citep{Popham_2001}. On the other hand, the Western model described the spectrum with single-temperature blackbody component and Comptonization arising from a multi-color disk component.
Additionally, a hybrid model \citep{hybrid} that combines the multi-color disk blackbody component in eastern model and the single-temperature blackbody component in western model along with a power-law component to represent  weak Comptonization was proposed later because both the Eastern and Western models encounter difficulties in the soft state of NS LMXBs. 
Therefore, it is common to use a double thermal model (a single-temperature blackbody component and a multi-color disk component) along with a Comptonization model as a starting point when doing X-ray spectrum analysis for NS LMXBs \citep{church2006,Coughenour_2018,Pandel_2008,wang_2019}.

In LMXBs, photons from the corona or BL can interact with the innermost region of the disk where they are absorbed and reprocessed by the material before being re-emitted. This is referred to as the ``reflection spectrum'' \citep{reflection}. The iron (Fe) K$\alpha$ line is one of the strongest emission lines within the reflection spectrum and contains information about the inner accretion flow within the shape of the line profile, which is broadened due to Doppler, special, and general relativistic effects. Through the modeling of reflection, additional information can be determined about the geometry of the accretion disk including the inner disk radius, inclination angle of the system \citep{Dauser_2010}, as well as the density and ionization state of the disk material (see \citealt{Ludlam24} for a recent review and additional references therein). 

The NS LMXB \source is a bright Z-source located near the Galactic Plane that has been observed extensively. 
BeppoSAX observations reported in \cite{Iaria_2006} determined that the spectra from 0.1 -- 30 keV can be well described by a single-temperature and Comptonized blackbody component from the NS surface with an additional power-law component required for higher energies. The blackbody radius $R_{BB}$ was found to be increasing from HB to FB. \cite{Seifina_2013} performed an analysis of  RTXE/PCA, Chandra/HETG and XMM-Newton data to study the spectral and timing properties of \source and found the photon index ($\Gamma$) of the high-energy Comptonized component remains relatively constant near 2 regardless of electron temperature and luminosity. \cite{Schulz_1993} analyzed EXOSAT data from 2--12 keV with a single-temperature blackbody component and a Comptonized blackbody component and found that the NS temperature increased from FB to HB. However, the Eastern model used in \cite{Schulz_1993} could not explain why the parameters change along the Z-track and evidence from the dipping class of LMXBs suggests that Comptonized emission originates from an extended accretion disc corona rather than a compact central Comptonizing region. Therefore, in \cite{church2006}, a single-temperature blackbody model plus Comptonization in an extended accretion disc corona (ADC) was used instead of Eastern model to describe the RXTE spectrum. The blackbody temperature $kT_{bb}$  decreased from HB to SA and a small increase from SA to FB, with the average temperature on FB still being the lower than NB and HB. A recent AstroSat spectral analysis \citep{chatt_2024} using the SXT and LAXPC found that the blackbody temperature $kT_{BB}$ increased from HB to HA then decreased along the NB and increases from SA to FB, while the $R_{BB}$ increased from HB to NB and decreased from NB to HB. The difference between this result and the existing literature is likely due to the fact that \cite{chatt_2024} modeled the data using a Comptonization and single-temperature blackbody without a component for the accretion disk.

In the broadband spectral analysis of BeppoSAX data from 0.1 -- 200 keV performed by \cite{Lavagetto_2004},
an Fe K$_{\alpha}$ emission line was observed and modeled by a Gaussian at $\sim$ 6.8 keV. In \cite{DA_2009}, the Fe K line was studied in the HB with XMM-Newton data by fitting it with a blackbody plus disk blackbody model in addition to a {\sc diskline} model to account for relativistic broadening of the line profile due to effects within the inner accretion disk. A blackbody temperature $kT_{bb} \sim$ 3 keV, disk temperature $T_{in} \sim$ 1.8 keV, inner disk radius $R_{in}$ of 13 gravitational radii ($R_g = GM/c^2$) and an inclination angle of 35$^\circ$ was reported \citep{DA_2009}. \cite{Cackett_2010} performed a systematic analysis of 10 NS LMXBs, including the XMM-Newton HB data of \source, using the double thermal model with a {\sc diskline} and found $kT_{bb}$ of 2 keV and inner disk radius $R_{in}$ in the range of $\sim 9-22\ R_g$ depending on modeling choices.

The bright nature and numerous existing studies makes \source a great candidate to compare the capabilities of \nustar serendipitous stray light observations with focused observations. In this paper, we examine the light curves and spectra of \source using 7 focused and 25 selected SL observations with the intention of demonstrating the utility of SL. The layout of the paper is as follows: \S2 presents the data selection and reduction, \S3 provides the spectral analysis and results, \S4 discusses the results and provides a comparison between datasets and the literature, and \S5 provides concluding remarks.

\begin{figure}[t!]
    \centering
    \includegraphics[width=0.46\textwidth, trim=1 0 0 0, clip]{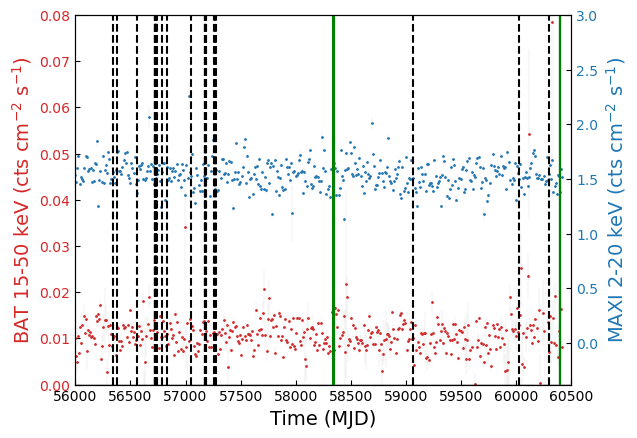}
    \caption{The long-term light curve of \source using MAXI 3--20 keV (blue) data and BAT 15--50 keV (red) data (when available). Vertical solid green lines indicate focused \nustar observations, whereas dashed black lines are SL observations. Note that some lines are thicker because multiple observations occurred close together.}
    \label{fig:long_lc}
\end{figure}

\section{Data reduction} \label{sec:data}
We consider data available on the archive through April 2024. The focused \nustar observations are taken in 2018 and 2024 while the serendipitous stray light observations occurred over the baseline of the mission with data from 2013, 2014, 2015, 2020, and 2023 (see Table~\ref{tab:staycats_info} in the Appendix for more details).
Figure~\ref{fig:long_lc} shows the long-term light curve for \source with the times of the SL and focused observations denoted. The SL observations are more numerous and cover a much wider range in time in comparison to the focused observations. Therefore, additional information may be gleaned when compared to the focused data. Both SL and focused observations were reduced with {\sc nustardas} v2.1.2.
The data were divided into their respective branch states based on inspection of the light curves. Additionally, light curves of each observation were searched for evidence of Type-I X-ray bursts, but do not find any in the observations considered. The reduction of focused and SL observations are discussed further in the following sections.

\subsection{Focused Observations}
\begin{figure*}[ht]
  \centering
  \begin{minipage}[b]{0.48\linewidth}
    \centering
    \includegraphics[width=\linewidth]{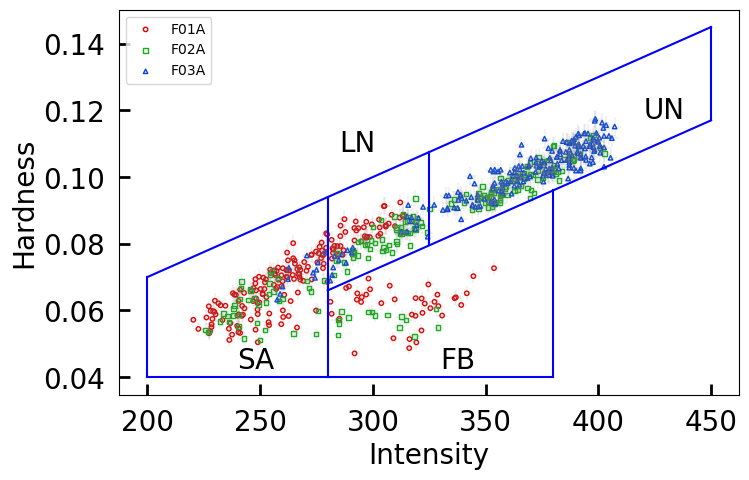}
  \end{minipage}\hfill
  \begin{minipage}[b]{0.48\linewidth}
    \centering
    \includegraphics[width=\linewidth]{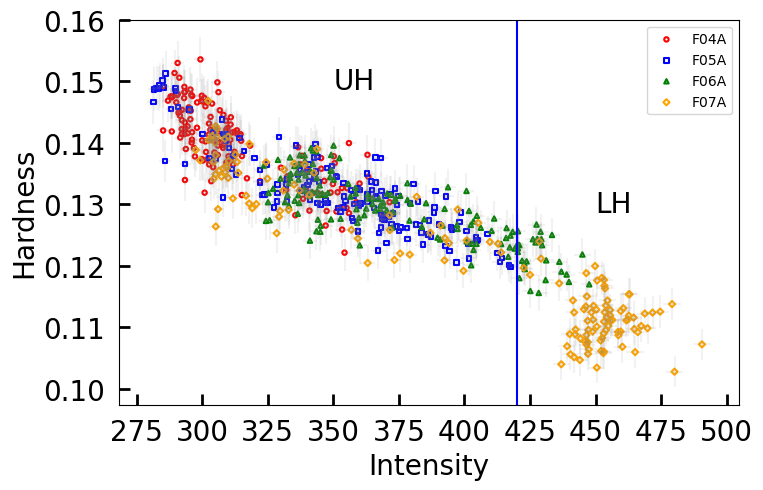}
  \end{minipage}
  \caption{HID for the focused observations. Only data from FPMA is shown for clarity. Left panel shows observations from 2018, right panel shows observations from 2024. The blue boxes/line indicate how the different branches are divided. The hardness ratio is defined as the hard energy band from 10--16 keV divided by the the soft energy band from 3--10 keV, while the total intensity is the 3--16 keV band. The HID is created with light curves binned at 128 s time bins. Naming scheme for the focused observations is provided in Table~\ref{tab:staycats_info}.}
  \label{fig:HI_focused}
\end{figure*}

A total of 7 focused observations are used for spectral analysis. Detailed information about the focused observations (including ObsID and obs\#) are given in Table~\ref{tab:staycats_info}. The cleaned, calibrated event files are produced by {\sc nupipeline 0.4.9}. {\sc nuproducts} was used to create light curves with 128 second time bins and extract spectra using a circular source region of $100^{\prime\prime}$ radius centered on the source and a background region of equal size sufficiently far from the source. Figure~\ref{fig:HI_focused} shows the HID for the focused data using 10--16 keV for hard energy band, 3--10 keV for soft energy band, and 3--16 keV for total intensity. The upper horizontal (UH), lower horizontal (LH), upper normal (UN), lower normal (LN), SA and FB are extracted based on the regions defined in the  HID  (Figure~\ref{fig:HI_focused}). To extract the spectra in the different branches, good time intervals (GTIs) are constructed to select the time bins that fall in each region. Spectra and light curves of each branches are then extracted with {\sc nuproducts} with the GTIs applied. Notably, all three focused observations in 2018 do not cover any HB at all. This is also the reason that we divided the NB into UN and LN. Since the HB is completely missing in these observations, we do not know if the ``upper normal'' branch actually contains a part of HA or not. Conversely, all four observations in 2024 contain data while the source was in the HB and the HID has shifted since 2018.
The 2024 observations were divided to the upper horizontal (UH) and lower horizontal (LH) branches. All spectra were optimally binned \citep{Kaastra_2016} via {\sc ftgrouppha} with a minimum number of 30 counts per bin to ensure the use of $\chi^2$ statistics. 

\subsection{Stray Light Observations}
The SL observations are extracted from StrayCats\footnote{https://nustarstraycats.github.io/straycats/} along with region files. A table of specific information about each SL observation is given in Table~\ref{tab:staycats_info} in the Appendix.
Since SL data has lower count rates compared to focused observations, additional selections must be made to ensure viable data. Similar to \cite{Brumback22}, we exclude observations with a region $\leq2\ \mathrm{cm}^2$ and where the source is contaminated by increased solar activity or another SL source (see Figure 3 in \citealt{Grefenstette_2021} for examples of different SL images). Out of 38 total available observations, 25 observations are considered further in this analysis after applying our selection criteria.

\begin{figure*}[ht]
\begin{center}
\includegraphics[width=0.8\textwidth, trim = 0 0 0 0, clip]{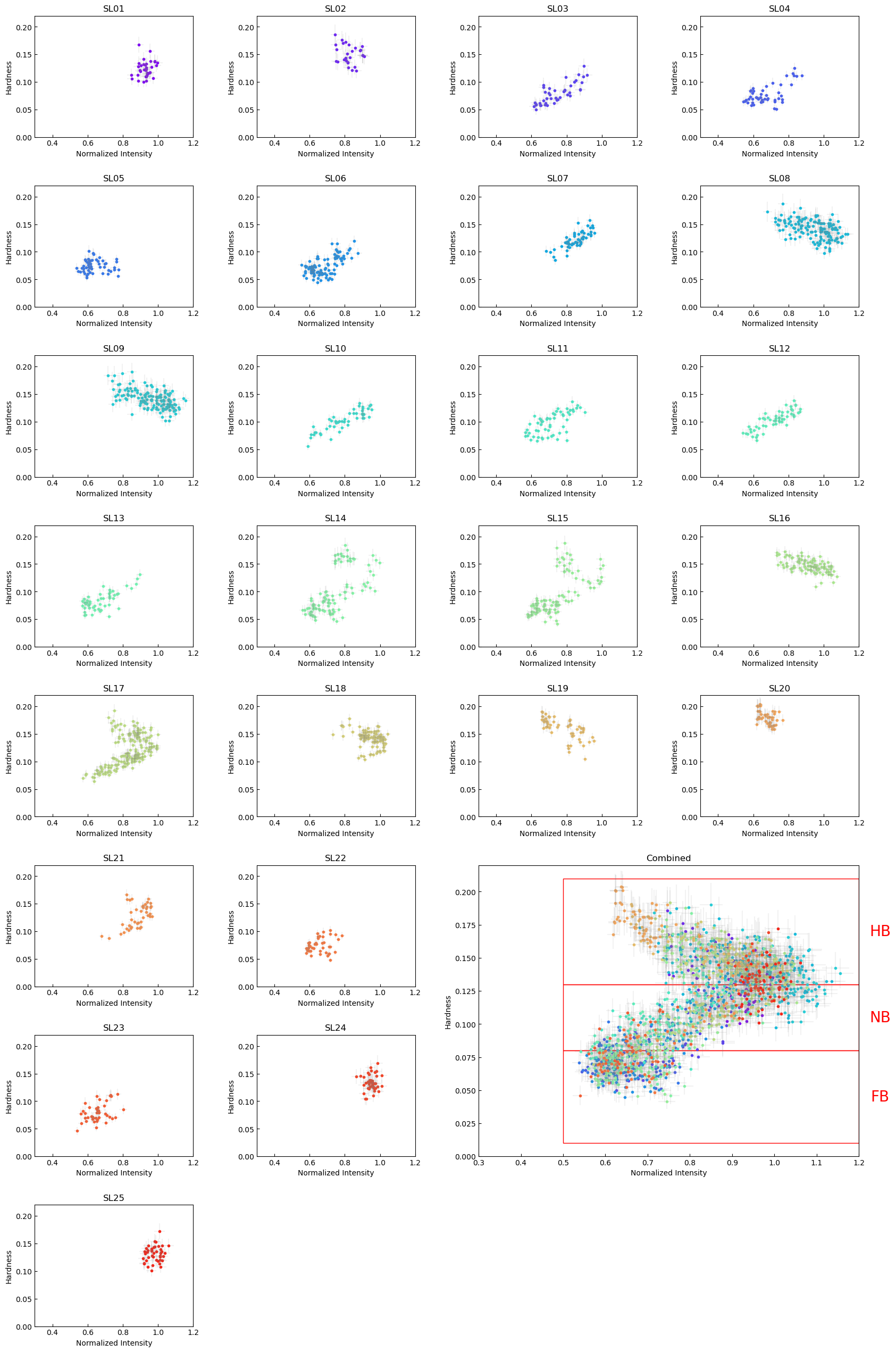}
\caption{HID for SL observations. The HID is constructed using the same energy bands as Figure~\ref{fig:HI_focused} with light curves using 512 s time bins. The combined plot on the bottom right shows the HID of all 25 observations concurrently. The red boxes indicate how the different branches are divided (HB, NB, and SA). Information (including naming scheme) for each SL observation is provided in Table~\ref{tab:staycats_info}.}
\label{fig:HI_stray}
\end{center}
\end{figure*}

Light curves and spectra are extracted with SL wrappers available in nustar-gen-utils\footnote{https://github.com/NuSTAR/nustar-gen-utils}. 
We plot the HID (Figure~\ref{fig:HI_stray}) for all observations using the same energy bands as with the focused data. In the HID, the total intensity is normalized by the SL area in order to directly compare between SL observations (which have various illuminating areas: see Table~\ref{tab:staycats_info}). We then define the GTIs for each state as indicated in the bottom right panel of Figure~\ref{fig:HI_stray} and spectra for each state using nustar-gen-utils. 
However, after extracting the data in the different branches, some spectra have $<5000$ total counts. We find these spectra to have too little data to return viable constraints in our analysis and do not consider them further. All remaining spectra were optimally binned  via {\sc ftgrouppha}.
It is important to note that unlike the focused observations, we only divided the HID for SL observation into 3 branches: HB, NB, and FB. This is again due to the lower count rate of SL. By only dividing the observations into 3 branches, we have higher photon counts for spectral analysis while still being able to compare the evolution of the system along the Z-track to the focused observations.

\begin{figure}[ht]
    \centering
    \includegraphics[width=0.48\textwidth]{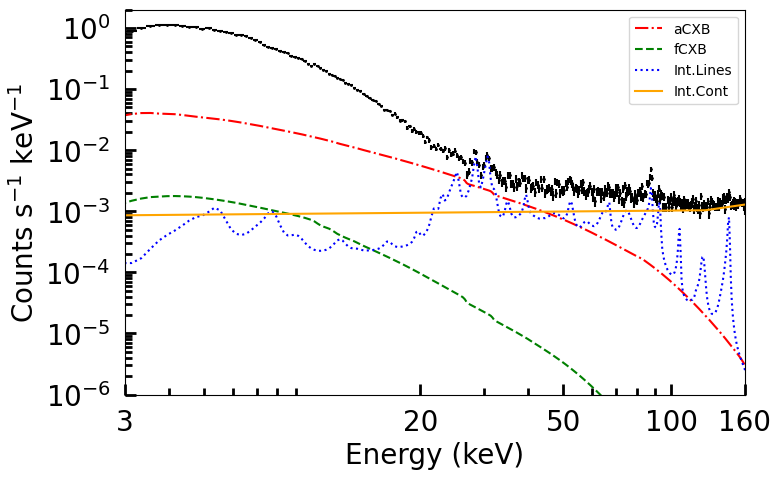} 
    \caption{An example SL spectrum (black) showing the background components modeled by {\sc nuskybgd}. Red dash-dot line is the aperture CXB, green dashed line is focused CXB, blue dotted line is the internal line emission, and orange solid line is the internal continuum.}
    \label{fig:background}
\end{figure}

Unlike with the focused observations, a background spectrum cannot be extracted directly from another region of the detector when handling SL data. There could be other faint sources at the level of the background or transmitted light within the extraction region that could alter the shape of our source spectrum if simply subtracted off \citep{straylight, Grefenstette_2021}. Thus, the background is modeled using {\sc nuskybgd} \citep{nuskybgd} to account for the astrophysical and instrumental background components. {\sc nuskybgd} is an adjustable background model that accounts for the cosmic X-ray background (CXB; containing a focused and an aperture component), solar background, and instrumental background (continuum and line emission). Instead of using a background region as described in the standard use of {\sc nuskybgd} for focused observations, we used the SL region directly and estimate the relative components therein for the background analysis. Since we removed the observations with high solar activity that are flagged in {\sc heasarc}, we turned off the solar flux that dominates below 5~keV. The {\sc nuskybgd} estimated value for the aperture CXB and focused CXB within the regions of interest were kept frozen for the rest of the analysis. To determine the instrumental background component contributions, we fit the SL spectrum in 80 -- 160 keV where there are no source counts from \source. The internal continuum of the instrumental background is fitted in 105 -- 160 keV range similar to \cite{Mastroserio_2022} and the internal lines were fit in 80 -- 160 keV range. Once the components for {\sc nuskybgd} are constrained, we freeze all the background components and continue with the source spectral analysis as discussed in the following section. An example background spectrum is shown in Figure~\ref{fig:background}.

\section{Spectral analysis \& results}
\begin{figure*}[ht]
    \centering
    \includegraphics[width=\textwidth]{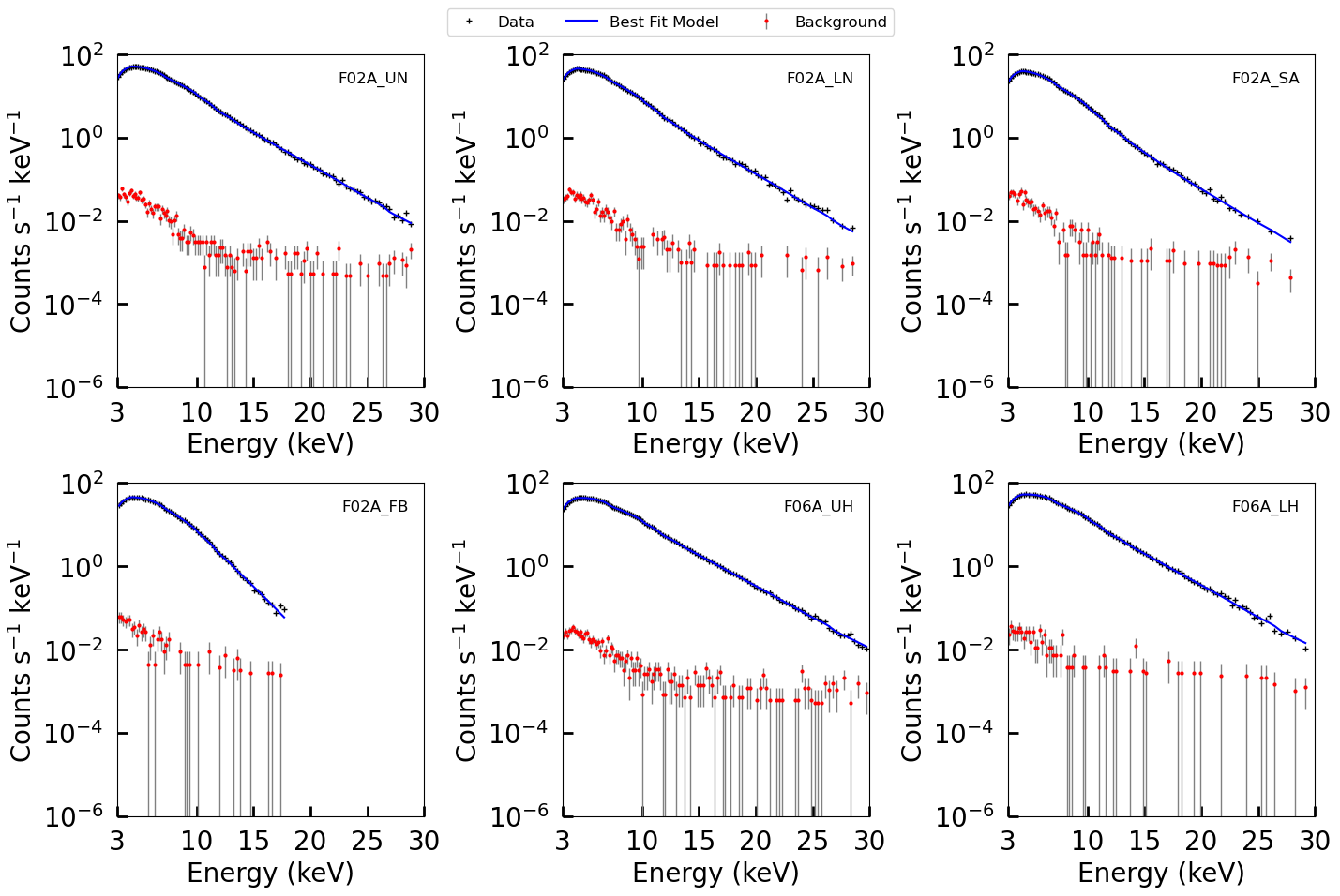} 
    \includegraphics[width=\textwidth]{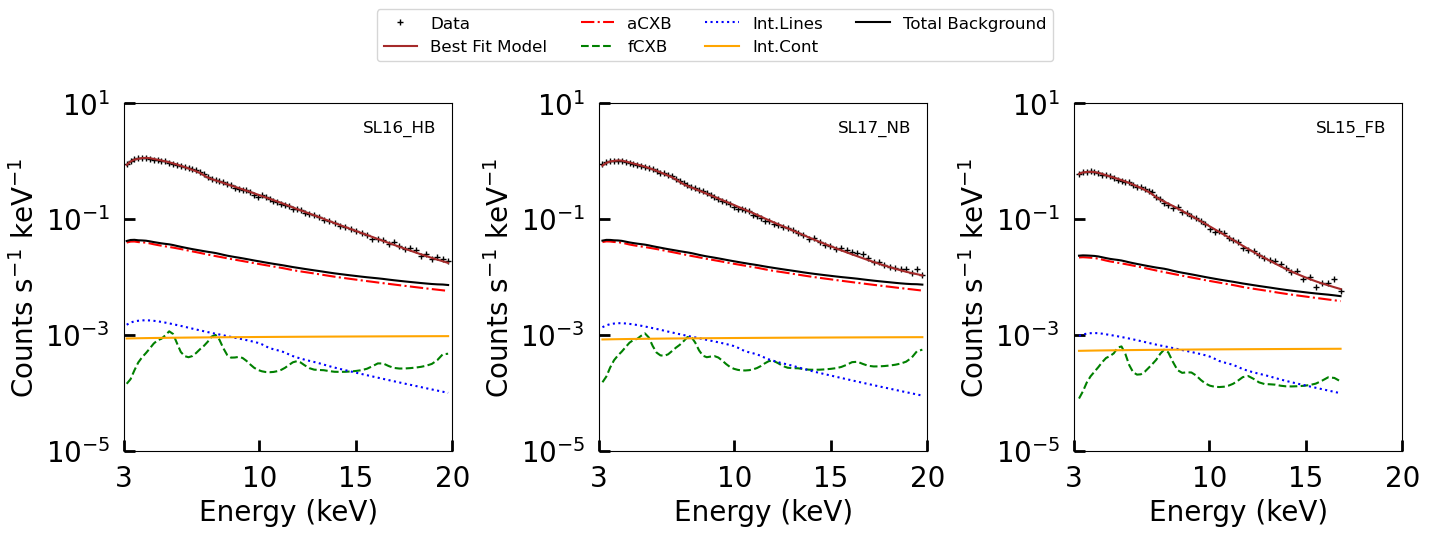} 
    \caption{Example spectra in various branches to illustrate the quality of the data for focused and SL observations. The top two rows show an example spectra for focused observations in each branch. The bottom row shows an example spectra of SL observations in each branch. The count rate per energy bin in SL data is more than 10 times lower than in the focused observations.}
    \label{fig:ld}
\end{figure*}

Spectral analysis was performed using {\sc xspec} version 12.13.0c \citep{arnaud96}. Spectra were considered from 3 keV to the energy where background dominates. The specific energy range for each observation is given in Table~\ref{tab:staycats_info} in the Appendix. Example spectra and background in each branch for focused and SL observations are given in Figure~\ref{fig:ld}. $\chi^2$ statistics were used when modeling the focused data while Cash statistics were used when fitting the SL data since the spectra have lower counts in comparison. Errors are reported at the 90\% confidence level throughout. 

\begin{figure*}[!t]
\begin{center}
\includegraphics[width=0.92\textwidth, trim = 1 1 0 0, clip]{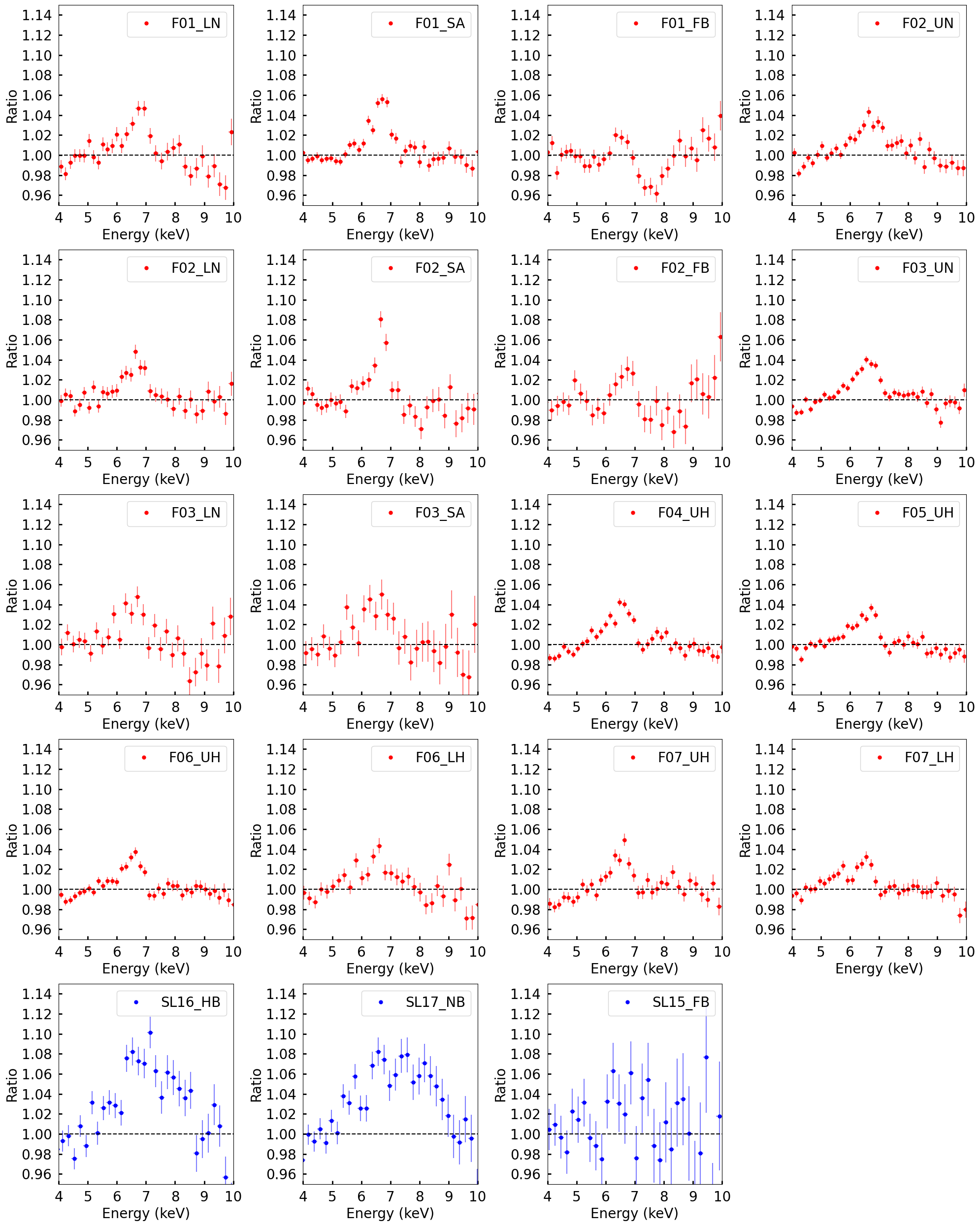}
\caption{Ratio of the data to the double thermal continuum model on the Fe line region for all focused observations (first four rows in red) and three select SL observation (final row in blue). The SL data have lower signal-to-noise than the focused data, as evidenced by the larger error bars on the data points. }
\label{fig:iron}
\end{center}
\end{figure*}

\subsection{Continuum model} 
The continuum was modeled using an absorbed double thermal model, i.e., {\sc tbabs * (bbody+diskbb)}.
We also tried to add a power-law or cut-off power-law to account for the weak Comptonization as often used in the hybrid model, but both led to nonphysical values for the fit  (e.g., power-law gives a low photon index $\Gamma \sim 0.5$ and cut-off power-law gives a low high energy cut off $\sim 0.7$).
We also tested different Comptonization models including {\sc nthcomp} \citep{1996MNRAS.283..193Z} and the convolution model {\sc thcomp} \citep{2020MNRAS.492.5234Z} for Comptonization from a blackbody component or disk component. These also led to nonphysical values with electron temperature $kT_e$ extremely high and covering fraction $cov_{\rm frac}$, which describes the fraction of seed photons that are scattered by the electron population, pegged at the limit of 1. This may be due to the lack of low-energy data available below 3 keV with \nustar. Regardless, there was no significant improvement in the fitting statistics. Furthermore, there currently are no self-consistent reflection models for a Comptonized continuum that are appropriate for NSs (see \citealt{Ludlam24} and references therein). {\sc RelxillCP} \citep{Garc_2014} was designed for BH systems and has a seed photon temperature too low for NSs, and {\sc Rfxconv} \citep{ref06} is an interpolation between two other models to create an approximate description of Comptonized reflection. Given that the focus in this analysis is to provide direct comparison between the focused and SL \nustar observations, we opt to focus on the single continuum description of the double thermal model applied to all spectra. A more detailed analysis of the continuum is outside the current scope of the paper.

\subsubsection{Continuum Fit Result}
Since the observations in both 2018 and 2024 occurred within approximately one week, a joint spectral fit was performed for each set of observations. Specifically, the neutral hydrogen column density ($N_H$) was tied together for all branches within each set of observations. This joint fit was conducted separately for the focused observations from 2018 and those from 2024. A constant was multiplied to the models to account for the differences between FPMA and FPMB. Table \ref{tab:Continuum_focused} shows the result of fitting the focused datasets.

\begin{table*}[t]
\begin{center}
\caption{Continuum fit for focused observations}
\label{tab:Continuum_focused}
\scriptsize
\begin{tabular}{ccccccccccc}
\toprule
&&{\sc tbabs} & \multicolumn{3}{c}{\sc bbody} & \multicolumn{3}{c}{\sc diskbb} &  \\
\cmidrule(lr){3-3}
\cmidrule(lr){4-6}
\cmidrule(lr){7-9}
obs\# & Branch & $N_{H}\ (\mathrm{10^{22}\ cm^{-2}})$ & $kT_{bb}(\mathrm{keV})$ & norm ($10^{-2}$) & $R_{bb}(\mathrm{km})$ & $T_{in}(\mathrm{keV})$ & norm & $R_{in}$ (km) & const & $\chi^2/dof$ \\
\midrule
F01 & LN & $7.35 \pm 0.05$ & $2.69 \pm 0.04$ & $2.23 \pm 0.12$ & $34 \pm 9$ & $1.73 \pm 0.01$ & $94 \pm 3$ & $20 \pm 5$ & $0.99 \pm 0.01$ & $355/192$ \\
& SA & * & $2.85 \pm 0.04$ & $0.93 \pm 0.04$ & $20 \pm 5$ & $1.73 \pm 0.01$ & $89 \pm 1$ & $19 \pm 5$ & $0.99 \pm 0.01$ & $994/205$ \\
& FB & * & $1.45 \pm 0.01$ & $14.59^{+0.13}_{-0.14}$ & $299 \pm 82$ & $0.80^{+0.03}_{-0.02}$ & $1257^{+224}_{-189}$ & $73 \pm 20$ & $0.99 \pm 0.01$ & $449/135$ \\
F02 & UN & * & $2.73 \pm 0.03$ & $3.45 \pm 0.11$ & $41 \pm 11$ & $1.81 \pm 0.01$ & $91 \pm 2$ & $19 \pm 5$ & $0.98 \pm 0.01$ & $566/220$ \\
& LN & * & $2.67 \pm 0.04$ & $2.28^{+0.11}_{-0.10}$ & $35 \pm 10$ & $1.72 \pm 0.01$ & $100 \pm 2$ & $20 \pm 5$ & $0.99 \pm 0.01$ & $408/201$ \\
& SA & * & $2.76 \pm 0.08$ & $0.81 \pm 0.07$ & $19 \pm 5$ & $1.69 \pm 0.01$ & $98 \pm 2$ & $20 \pm 5$ & $1.00 \pm 0.01$ & $524/179$ \\
& FB & * & $1.43 \pm 0.01$ & $13.77^{+0.27}_{-0.31}$ & $299 \pm 82$ & $0.86^{+0.05}_{-0.04}$ & $905^{+270}_{-205}$ & $61 \pm 19$ & $0.99 \pm 0.01$ & $211/128$ \\
F03 & UN & * & $2.76 \pm 0.02$ & $3.73 \pm 0.08$ & $42 \pm 11$ & $1.83 \pm 0.01$ & $88 \pm 1$ & $19 \pm 5$ & $0.98 \pm 0.01$ & $983/235$ \\
& LN & * & $2.68 \pm 0.06$ & $2.31^{+0.17}_{-0.16}$ & $35 \pm 10$ & $1.73 \pm 0.02$ & $98^{+4}_{-3}$ & $20 \pm 5$ & $1.03 \pm 0.01$ & $332/176$ \\
& SA & * & $2.64 \pm 0.10$ & $1.35^{+0.18}_{-0.16}$ & $27 \pm  8$ & $1.70 \pm 0.02$ & $100 \pm 4$ & $20 \pm 5$ & $1.00 \pm 0.01$ & $241/162$ \\
F04 & UH & $6.93 \pm 0.05$ & $2.82 \pm 0.02$ & $5.17 \pm 0.11$ & $47 \pm 13$ & $1.94 \pm 0.01$ & $47 \pm 1$ & $14 \pm 3$ & $0.99 \pm 0.01$ & $1117/239$ \\
F05 & UH & * & $2.82 \pm 0.02$ & $5.16 \pm 0.11$ & $47 \pm 13$ & $1.95 \pm 0.01$ & $55 \pm 1$ & $15 \pm 3$ & $0.99 \pm 0.01$ & $796/239$ \\
F06 & UH & * & $2.77 \pm 0.02$ & $5.48 \pm 0.12$ & $50 \pm 14$ & $1.89 \pm 0.01$ & $63^{+2}_{-1}$ & $17 \pm 5$ & $1.02 \pm 0.01$ & $711/237$ \\
& LH & * & $2.85 \pm 0.05$ & $4.97^{+0.33}_{-0.32}$ & $45 \pm 13$ & $1.97 \pm 0.03$ & $68 \pm 3$ & $17 \pm 5$ & $1.00 \pm 0.01$ & $272/188$ \\
F07 & UH & * & $2.78 \pm 0.02$ & $5.36 \pm 0.15$ & $49 \pm 13$ & $1.88 \pm 0.02$ & $59 \pm 2$ & $15 \pm 5$ & $0.99 \pm 0.01$ & $621/226$ \\
& LH & * & $2.82 \pm 0.03$ & $4.70 \pm 0.17$ & $45 \pm 12$ & $1.94 \pm 0.01$ & $80 \pm 2$ & $19 \pm 5$ & $0.98 \pm 0.01$ & $437/225$ \\
\hline
\multicolumn{3}{l}{$^* = $ Tied to same value above}
\end{tabular}
\end{center}
\smallskip
Note. -- $R_{BB}$ and $R_{in}$ are calculated with a color correction factor of $f_{cor}=1.7$ \citep{Shimura_1995} and a source distance of $11 \pm 3$~kpc \citep{Fender_2000}. An inclination angle of 34.08$^\circ$ is used for $R_{in}$, which is the average value determined from reflection modeling in Table~\ref{tab:reflection_focused}.
\end{table*}

Because illumination of the detectors is a geometric effect depending on the orientation of the telescope and off axis angle of the source, the majority of the SL data only exist on one FPM for a given observation ID, thus we are unable to analyze both FPMA and FPMB together as would occur for focused data. As a result, each SL observation is considered individually. We fix the column density when modeling the SL spectra to the average value from the focused data. When left as a free parameter, inferred values varied over a large range between spectra with some outside values previously reported in literature ($N_{H}\sim4-11\times10^{22}\ \rm cm^{-2}$: \citealt{Lavagetto_2004,Iaria_2006,DA_2009,Cackett_2010,chatt_2024}). We discuss the potential impact of $N_{H}$ on the results in \S\ref{sec:4p1}.
Table \ref{tab:continuum_SL_example} shows the result of an example SL observation for each branch. Full result can be found in Table \ref{continuum_full}.

\begin{table*}[ht]
\begin{center}
\caption{Continuum fit for stray light observations with highest photon counts in each branch}
\label{tab:continuum_SL_example}
\begin{tabular}{cccccccccc}
\toprule
& & \multicolumn{3}{c}{\sc bbody} & \multicolumn{3}{c}{\sc diskbb} &  \\
\cmidrule(lr){3-5}
\cmidrule(lr){6-8}
obs\# & Branch & $kT_{bb}\ \rm (keV)$ & norm ($10^{-2}$) & $R_{BB}\ \rm (km)$ & $T_{in}\ \rm (keV)$ & norm & $R_{in}$ (km) & C/dof \\
\midrule
SL16 & HB & $2.17^{+0.03}_{-0.03}$ & $11.2^{+0.4}_{-0.4}$ & $117^{+32}_{-32}$ & $1.22^{+0.03}_{-0.03}$ & $453^{+51}_{-45}$ & $44^{+12}_{-12}$ & $201/70$\\
SL17 & NB & $1.86^{+0.03}_{-0.03}$ & $9.74^{+0.5}_{-0.5}$ & $148^{+41}_{-41}$ & $1.12^{+0.04}_{-0.03}$ & $682^{+99}_{-85}$ & $53^{+15}_{-15}$ & $212/69$\\
SL15 & FB & $1.39^{+0.02}_{-0.02}$ & $12.45^{+0.6}_{-0.6}$ & $301^{+83}_{-83}$ & $0.7^{+0.06}_{-0.05}$ & $5351^{+3380}_{-2001}$ & $150^{+63}_{-49}$ & $62/53$\\
\bottomrule
\end{tabular}
\end{center}
\vspace{2mm}

Note. -- The column density $N_H\ (10^{22}\ \mathrm{cm}^{-2})$ is fixed at 6.39, which is the average value found in Table~\ref{tab:reflection_focused}. $R_{BB}$ and $R_{in}$ are calculated using the same values for $f_{cor}$, distance, and inclination as Table~\ref{tab:Continuum_focused}.

\end{table*}

\subsection{Reflection model}
An Fe K emission line peaking between 6.4--6.97 keV is found in all observations (see Figure~\ref{fig:iron}), which indicative of reflection from the accretion disk. We applied the self-consistent relativistic reflection model {\sc relxillNS} \citep{García_2022} for thermal illumination of the accretion disk. It uses an empirical power-law emissivity with a blackbody component as the primary source. Since a blackbody with temperature $kT_{bb}$ is already included in the model as the incident spectrum, we removed the {\sc BBODY} component from the overall model. The emissivity profile of the reflection component follows a broken power-law, where the inner disk is characterized by an index $q_1$, and the outer disk beyond the break radius $R_{\text{break}}$ follows a flatter profile with an index $q_2$. This transition from $q_1$ to $q_2$ at $R_{\text{break}}$ accounts for changes in the illumination pattern across the disk, typically due to relativistic effects and geometric factors near the compact object. However, \cite{emissivity} showed that a single emissivity index $q$ near 3 is sufficient when the disk is illuminated by different sources including hotspots on NS surfaces, bands of emission, and the entire hot, spherical star surface, thus we use a single emissivity index ($q_1=q_2$) fixed at 3. We note that this choice does not impact our subsequent analysis as the results are consistent when the emissivity index is allowed to be free. However, for some SL observations, allowing the emissivity index to be free will cause the result to be less constrained and give larger errors. Therefore, we leave $q$ fixed for consistency between the analysis of the focused and stray light data. The break radius is fixed at 500 \rg (an obsolete parameter when using a single emissivity profile) and the outer disk radius is fixed at 1000 \rg while allowing the inner disk radius to be free (in units of \risco). The dimensionless spin $a$, which defines the location of the innermost stable circular orbit $R_{ISCO}$, is expected to be smaller than 0.3 for Galactic NS LMXBs \citep{Galloway_2008} and the difference between $a=0$ and $a=0.3$ in \risco is small (about 1 $R_g$), so we fixed $a$ at 0. The intrinsic parameters of the disk including ionization $\log(\xi)$ and iron abundance ($A_{Fe}$) are free to vary while the disk density ($\log(n_e/\rm cm^{-3})$) is fixed at the maximum allowed value in the model (which is 19) since the accretion disk around NSs are expected to have a higher disk density \citep{Frank_2002,Shakura_1973,Enzo_2023,Ludlam24}. The redshift, $z$, is fixed at 0 since \source is a galactic source.

\subsubsection{Reflection Fit Results}
\begin{figure*}[ht]
  \centering
  \begin{minipage}[b]{0.48\linewidth}
    \centering
    \includegraphics[width=\linewidth]{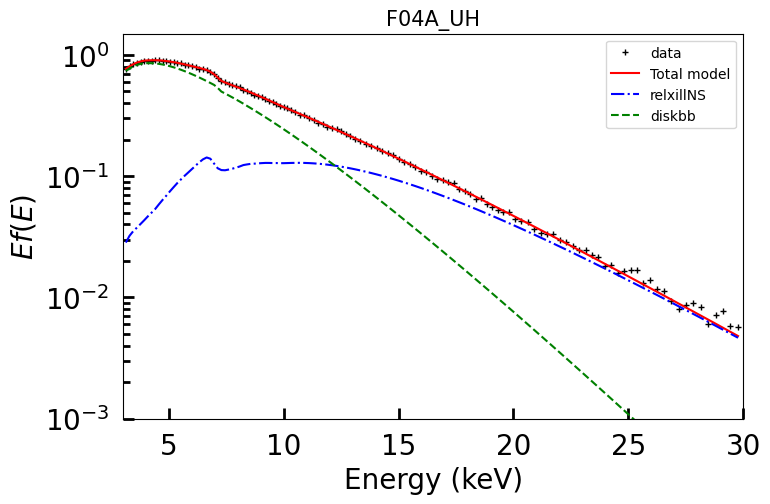}
  \end{minipage}\hfill
  \begin{minipage}[b]{0.48\linewidth}
    \centering
    \includegraphics[width=\linewidth]{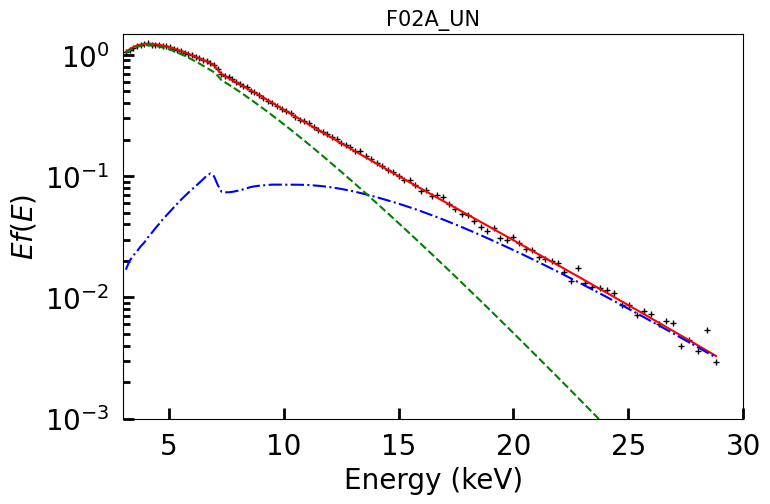}
  \end{minipage}
  \vfill
  \begin{minipage}[b]{0.48\linewidth}
    \centering
    \includegraphics[width=\linewidth]{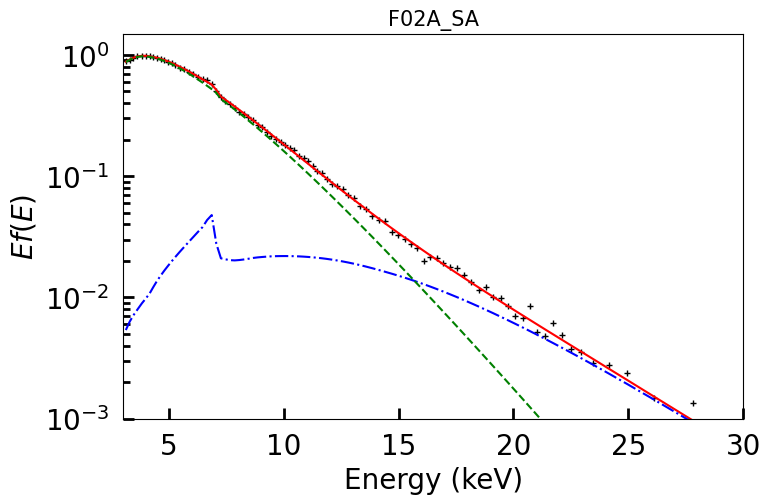}
  \end{minipage}\hfill
  \begin{minipage}[b]{0.48\linewidth}
    \centering
    \includegraphics[width=\linewidth]{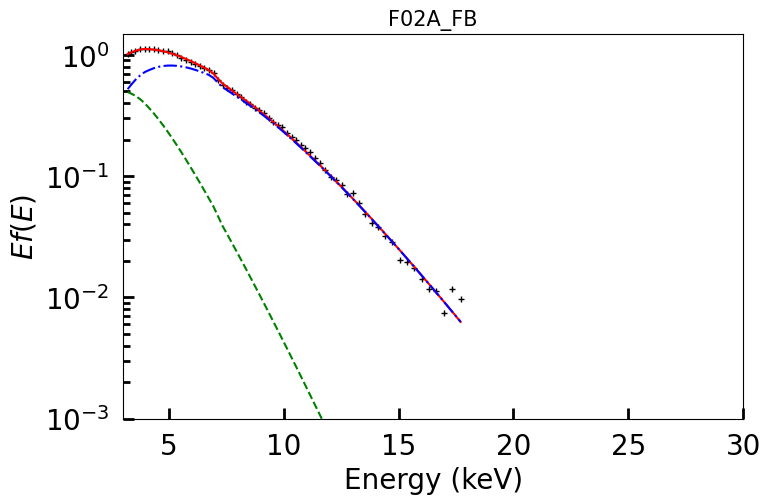}
  \end{minipage}
  \caption{Example of unfolded spectra for focused observations in each branch to show the contribution of the reflection (blue dot-dashed line) and disk (green dashed line) components.}
  \label{fig:unfolded_focused}
\end{figure*}

\begin{figure*}[ht]
    \centering
    \includegraphics[width=\textwidth]{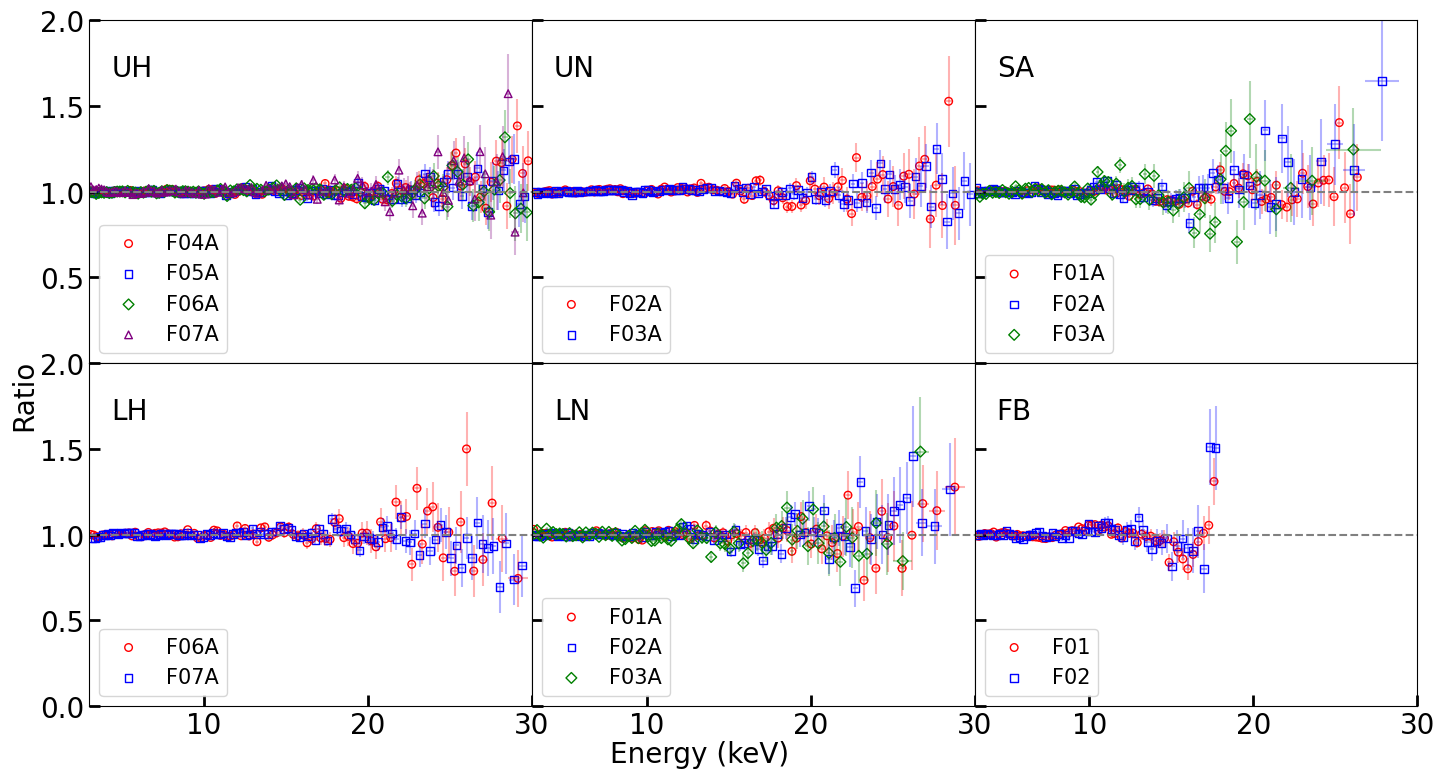}
    \\[1em] 
    \includegraphics[width=\textwidth]{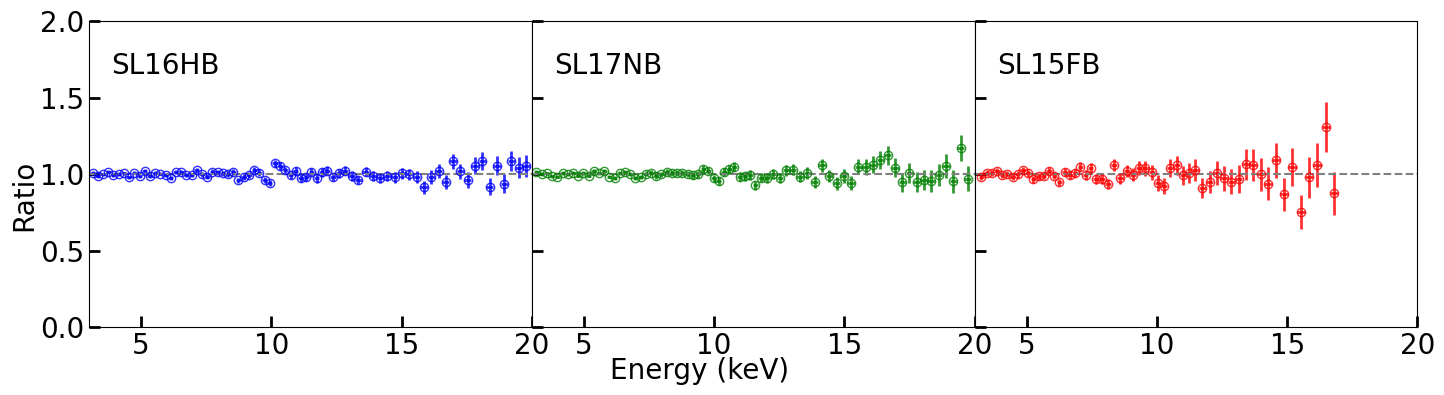}
    \caption{Ratio of the data to the overall model including reflection in focused and select SL observations to demonstrate the goodness of fit. The top six panels show the ratio plot for all focused observations while the lower three panels show an example of the ratio plot for each branch of the SL data.}
    \label{fig:ratio}
\end{figure*}

Similar to the continuum modeling, we did a joint fit for each set of focused observations whereas the SL data were modeled individually. The hydrogen column density ($N_H$), iron abundance ($A_{Fe}$), and inclination angle ($i$) are tied together when jointly modeling the focused observations. The results for the focused data are given in Table \ref{tab:reflection_focused}. An example of the unfolded spectrum and ratio plots for each branch is shown in Figure \ref{fig:unfolded_focused}. The full ratio plots are shown in Figure \ref{fig:ratio} after modeling the reflection. Table~\ref{tab:reflection_SL_example} provides the results of a select SL observation in each branch and the ratio of the data to the model can be seen in the last row of Figure \ref{fig:ratio}. The full results of all SL observations can be found in Table~\ref{tab:reflection_all}.

\begin{table*}[ht]
\begin{center}    
\caption{Reflection fit for focused observations}
\label{tab:reflection_focused}
\scriptsize
\begin{tabular}{cccccccccccc}

\toprule
&& \multicolumn{6}{c}{\sc relxillNS} & \multicolumn{2}{c}{\sc diskbb} &  \\
\cmidrule(lr){3-8}
\cmidrule(lr){9-10}
obs\# & Branch  & $R_{in}$ (\risco) & $kT_{bb}$ (keV) & $\log(\xi)$ & $A_{Fe}$ & $f_{refl}$&$\rm norm\ (10^{-3})$&$T_{in}$ (keV)&norm &const& $\chi^2/dof$ \\
\midrule
F01 & LN&${1.48}^{+0.07}_{-0.12}$&${2.78}^{+0.09}_{-0.01}$&${1.78}^{+0.02}_{-0.13}$&${1.95}^{+0.07}_{-0.20}$&${2.0} \pm 0.1$&${1.1} \pm 0.1$&${1.79}^{+0.02}_{-0.01}$&${81}^{+1}_{-3}$&${0.99} \pm 0.01$& 3862/1792\\
& SA &${1.19}^{+0.11}_{-0.04}$&${2.78}^{+0.07}_{-0.04}$&${2.00} \pm 0.02$&${3.7}^{+0.2}_{-0.3}$&${1.5} \pm 0.1$&${0.6} \pm 0.1$&${1.75} \pm 0.01$&${82} \pm 1$&${0.99} \pm 0.01$\\
& FB &${1.48}^{+0.04}_{-0.23}$&${1.50} \pm 0.01$&${2.00}^{+0.01}_{-0.03}$&${0.70}^{+0.03}_{-0.01}$&${7.6}^{+0.2}_{-0.7}$&${2.6}^{+0.3}_{-0.1}$&${1.25}^{+0.01}_{-0.04}$&${222}^{+21}_{-4}$&${0.99} \pm 0.01$\\
F02 & UN&${1.17}^{+0.01}_{-0.05}$&${3.07}^{+0.05}_{-0.01}$&${2.31}^{+0.08}_{-0.10}$&${0.71}^{+0.06}_{-0.04}$&${4.6}^{+0.1}_{-0.4}$&${0.8} \pm 0.1$&${1.91} \pm 0.01$&${74}^{+1}_{-2}$&${0.98} \pm 0.01$ \\
& LN &${1.4}^{+0.3}_{-0.2}$&${2.88}^{+0.10}_{-0.01}$&${2.03}^{+0.02}_{-0.01}$&${0.53}^{+0.08}_{-0.01}$&${3.6}^{+0.3}_{-0.2}$&${0.7} \pm 0.1$&${1.80} \pm 0.01$&${84}^{+1}_{-2}$&${0.99} \pm 0.01$\\
& SA &${1.41}^{+0.19}_{-0.04}$&${2.78}^{+0.01}_{-0.12}$&${2.01}^{+0.11}_{-0.05}$&${3.53}^{+0.43}_{-0.02}$&${2.6}^{+0.4}_{-0.1}$&${0.4} \pm 0.1$&${1.72} \pm 0.01$&${88} \pm 1$&${1.00} \pm 0.01$\\
& FB&${1.34}^{+0.09}_{-0.14}$&${1.43} \pm 0.01$&${1.11}^{+0.08}_{-0.09}$&${1.08}^{+0.03}_{-0.15}$&${0.6} \pm 0.1$&${12.0}^{+0.1}_{-0.3}$&${0.94}^{+0.02}_{-0.01}$&${604}^{+4}_{-56}$&${0.99} \pm 0.01$ \\
F03 & UN &${1.12}^{+0.05}_{-0.03}$&${3.04}^{+0.05}_{-0.01}$&${2.08}^{+0.11}_{-0.06}$&${0.51}^{+0.09}_{-0.01}$&${4.9}^{+0.1}_{-0.5}$&${0.8} \pm 0.1$&${1.93} \pm 0.01$&${71} \pm 1$&${0.98} \pm 0.01$\\
& LN&${1.22}^{+0.03}_{-0.17}$&${3.02}^{+0.02}_{-0.09}$&${2.10}^{+0.07}_{-0.08}$&${0.61}^{+0.04}_{-0.01}$&${5.8}^{+0.5}_{-0.1}$&${0.4} \pm 0.1$&${1.82} \pm 0.01$&${79}^{+2}_{-1}$&${1.03} \pm 0.01$ \\
& SA&${2.23}^{+0.08}_{-0.31}$&${2.84}^{+0.09}_{-0.10}$&${2.04}^{+0.02}_{-0.18}$&${0.71}^{+0.03}_{-0.04}$&${4.8}^{+0.4}_{-0.5}$&${0.3} \pm 0.1$&${1.75} \pm 0.01$&${86} \pm 2$&${0.99} \pm 0.01$\\

F04 & UH &${1.49}^{+0.11}_{-0.08}$&${3.24}^{+0.02}_{-0.01}$&${2.57}^{+0.01}_{-0.02}$&${0.60}^{+0.03}_{-0.02}$&${3.5}^{+0.3}_{-0.2}$&${1.20}^{+0.04}_{-0.09}$&${2.11}\pm0.01$&${35}\pm1$&${0.99}\pm0.01$&1915/1328\\
F05 & UH &${1.44}^{+0.08}_{-0.10}$&${3.38}^{+0.02}_{-0.01}$&${2.58}\pm0.02$&${0.59}\pm0.01$&${7.6}^{+0.7}_{-0.4}$&${0.60}^{+0.03}_{-0.05}$&${2.14}\pm0.01$&${39}\pm1$&${0.99}\pm0.01$\\
F06 & UH &${1.48}^{+0.07}_{-0.15}$&${3.32}\pm0.01$&${2.52}^{+0.01}_{-0.02}$&${0.56}^{+0.04}_{-0.01}$&${5.8}^{+0.2}_{-0.5}$&${0.77}^{+0.06}_{-0.03}$&${2.10}\pm0.01$&${44}\pm1$&${1.01}\pm0.01$ \\
& LH &${1.3}^{+0.2}_{-0.2}$&${3.45}^{+0.04}_{-0.05}$&${2.55}\pm0.04$&${0.76}^{+0.09}_{-0.04}$&${9.7}^{+1.1}_{-1.0}$&${0.46}^{+0.06}_{-0.04}$&${2.14}\pm0.02$&${50}\pm2$&${1.00}\pm0.01$\\
F07 & UH &${1.06}^{+0.03}_{-0.05}$&${3.33}^{+0.06}_{-0.02}$&${2.48}\pm0.02$&${0.77}\pm0.03$&${3.2}^{+0.8}_{-0.3}$&${1.22}^{+0.07}_{-0.22}$&${2.13}^{+0.02}_{-0.01}$&${38}\pm1$&${0.99}\pm0.01$\\
& LH &${1.02}^{+0.08}_{-0.02}$&${3.28}^{+0.04}_{-0.02}$&${2.38}^{+0.04}_{-0.06}$&${0.98}^{+0.08}_{-0.09}$&${3.4}^{+0.4}_{-0.3}$&${1.14}^{+0.07}_{-0.13}$&${2.09}\pm0.01$&${60}\pm1$&${0.98}\pm0.01$\\

\bottomrule
\hline

\end{tabular}
\end{center}
\vspace{2mm}
Note. -- The disk density $\log(n_e/\rm cm^{-3})$ is fixed at 19, spin is fixed at $a=0$, and emissivity index is fixed at $q=3$. Data from the same year were modeled concurrently. For the 2018 dataset (F01--F03), the column density was found to be $N_H\ (10^{22}\ \rm cm^{-2})={6.78}^{+0.03}_{-0.09}$ and inclination of $i\ (^\circ)={38.36}^{+0.06}_{-1.43}$. For the 2024 dataset (F04--F07), the column density was found to be $N_H\ (10^{22}\ \rm cm^{-2})={6.07}^{+0.01}_{-0.04}$ and inclination of $i\ (^\circ) = {29.8}^{+0.3}_{-0.3}$.
\end{table*}

\begin{table*}[ht]
\begin{center}
    
\caption{Reflection fit for select SL observations with highest photon counts in each branch}
\label{tab:reflection_SL_example}
\begin{tabular}{ccccccccccccc}
\toprule
\multicolumn{2}{c}{} & \multicolumn{6}{c}{\sc relxillNS} & \multicolumn{2}{c}{\sc diskbb} &  \\
\cmidrule(lr){3-8}
\cmidrule(lr){9-10}
obs\# & Branch  & $R_{in}$ (\risco) & $kT_{bb}$ (keV) & $\log(\xi)$ & $A_{Fe}$ & $f_{refl}$&$\rm norm\ (10^{-3})$&$kT_{in}$(keV)&norm& $C/dof$ \\
\midrule
SL16 & HB&${8.4}_{-6.6}^{+1.2}$ & ${2.22}^{+0.10}_{-0.06}$ & ${3.9}^{+0.7}_{-2.2}$ & ${9.7}^{+0.3}_{-9.1}$ & ${0.34}^{+1.06}_{-0.21}$ & ${9.3}^{+1.9}_{-4.2}$ & ${1.12}^{+0.10}_{-0.07}$ & ${586}^{+172}_{-202}$ & $93/66$ \\
SL17 & NB& ${2.1}_{-0.5}^{+7.3}$ & ${2.14}^{+0.16}_{-0.05}$ & ${4.0}^{+0.6}_{-0.2}$ & ${3.1}^{+1.7}_{-1.5}$ & ${16}^{+2}_{-10}$ & ${0.6}^{+2.2}_{-0.1}$ & ${0.97}^{+0.06}_{-0.08}$ & ${1179}^{+127}_{-286}$ & $78/65$\\
SL15& FB &${1.9}_{-0.5}^{+6.8}$ & ${1.45}^{+0.05}_{-0.06}$ & ${2.5}^{+0.8}_{-0.6}$ & ${9.6}^{+0.1}_{-9.1}$ & ${0.34}^{+2.41}_{-0.21}$ & ${10.5}^{+1.5}_{-6.1}$ & ${0.81}^{+0.12}_{-0.09}$ & ${2320}^{+2031}_{-1157}$ & $49/50$\\
\bottomrule
\hline
\end{tabular}
\end{center}
\vspace{2mm}

Note.-- The disk density $\log(n_e)$ is fixed at 19 cm$^{-3}$, spin is fixed at $a=0$, and emissivity index is fixed at $q=3$. The column density $N_H\ (10^{22}\ \rm cm^{-2})$ and inclination is fixed at the average values from the focused spectral analysis of ${6.39}$ and $34.08^\circ$, respectively. Results for all SL observations are provided in Table~\ref{tab:reflection_all}.

\end{table*}

\section{Discussion}
We have performed a spectral analysis on \nustar data of \source using 7 focused observations and 25 SL observations. The The data were divided up into different spectral states spanning the HB to the FB. The continuum was described using a double thermal model and the remaining reprocessed emission from the accretion disk was fit using \relxillns.
In this section, we compare our results between focused and SL data, as well as with the existing literature, and discuss any inconsistencies.

\subsection{Continuum model}\label{sec:4p1}
The results of continuum modeling for focused and SL observations show similar trend for blackbody temperature $kT_{BB}$ and disk temperature $T_{in}$. From the fit result, we calculated the blackbody radius $R_{BB}$ and the inner disk radius $R_{in}$. The blackbody radius $R_{BB}$ is calculated using standard blackbody luminosity formula with a color correction factor $f_{cor}=1.7$ \citep{Shimura_1995}. The luminosity is obtained from the {\sc bbody} normalization assuming a distance of $11\pm3$~kpc \citep{Fender_2000}. The luminosity of the blackbody in focused observations spans the range of $0.1 \times 10^{38}$~\lumcgs -- $1.8 \times 10^{38}$~\lumcgs. The luminosity of the blackbody in SL observations spans $0.9 \times 10^{38}$~\lumcgs -- $1.5 \times 10^{38}$~\lumcgs. The inner disk radius $R_{in}$ is obtained from the {\sc diskbb} normalization with a color correction factor $f_{cor}=1.7$ using the average inclination angle obtained from the reflection fitting of the focused observations in Table~\ref{tab:reflection_focused}. 
For both focused and SL observations, the blackbody temperature $kT_{BB}$ and disk temperature $T_{in}$ are at the lowest in FB, then remain almost the same from NB to HB. The blackbody radius $R_{BB}$ is at the highest in FB, then it decreases from FB to NB and increases from NB to HB, which is consistent with the result from \citep{church2006}. The inner disk radius $R_{in}$ obtained from the continuum model agrees with the result in the reflection model in focused observations except for FB where the {\sc diskbb} normalization is poorly constrained. 
The $R_{in}$ in SL observations are all poorly constrained in the reflection model. Therefore, we do not compare the $R_{in}$ between continuum and reflection model for SL. The reason that $R_{in}$ is poorly constrained in the reflection model for SL observations is discussed further in \S\ref{sec:4p2}. 
Despite that, the blackbody temperature $kT_{bb}$ remains nearly constant in the NB and HB for both focused and SL observations. 

\begin{figure*}[t]
    \centering
    
    (a)\\
    \includegraphics[width=0.8\textwidth]{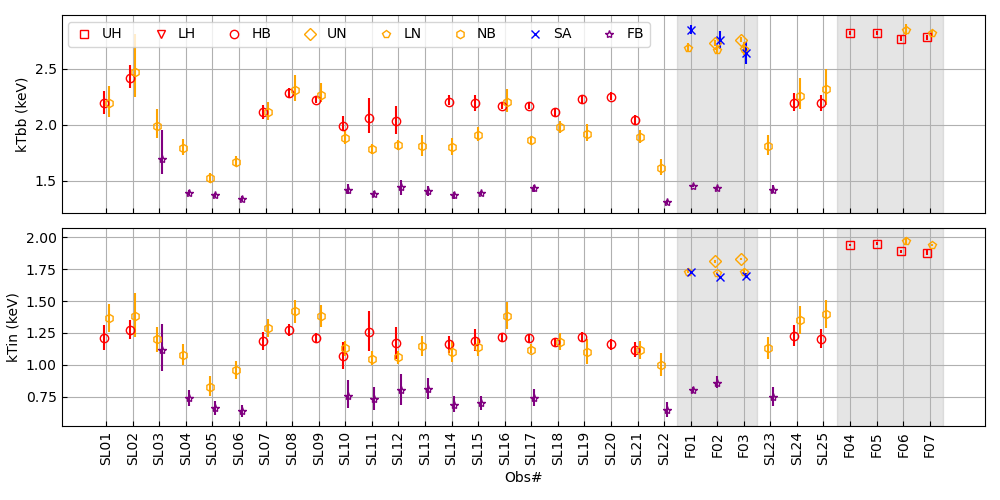} 
    
    \vspace{0.1cm} 
    (b)\\
    \includegraphics[width=0.8\textwidth]{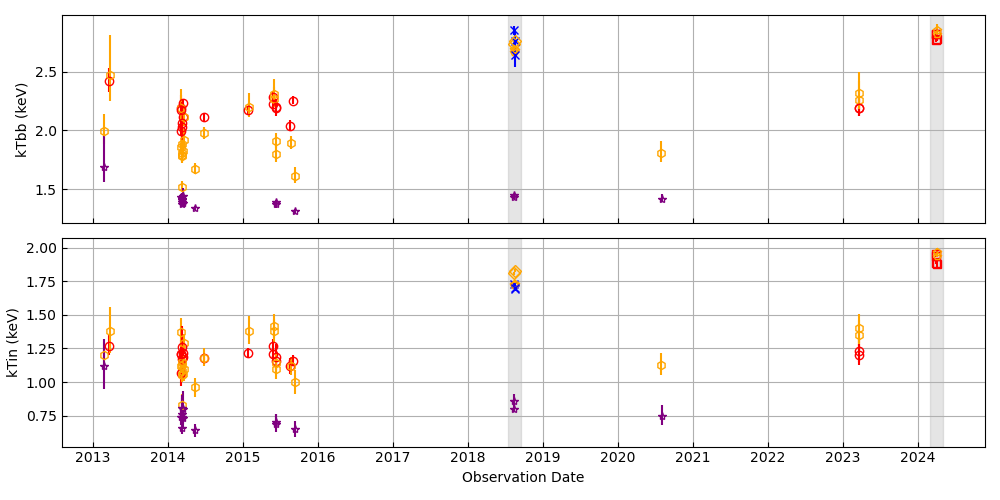}
    
    \caption{Variation in blackbody temperature $kT_{bb}$ and disk temperature $T_{in}$ over time for observations from 2013 to 2024. The two upper panels below (a) display the temperatures for each observation, arranged sequentially along the x-axis with equal spacing between points to emphasize individual observation order. The two lower panels below (b) present the same temperature variations but the observations are plotted versus time; illustrating the actual temporal distribution of the data. Colors and marker shapes represent different branches: the horizontal branch (HB) is in red, normal branch (NB) in orange, soft apex (SA) in blue, and flaring branch (FB) in purple. Within the HB and NB, the upper-horizontal (UH) is denoted by a square, lower-horizontal (LH) by downward pointing triangle, upper-normal (UN) by a diamond, and lower-normal (LN) sections by a pentagon. Observations from stray light (SL) in the HB and NB are marked by a circle and hexagon. For additional distinction between observation types, shaded regions indicate focused observations.}
    \label{fig:parameters}
\end{figure*}

The focused observations show a higher blackbody and disk temperature in the NB and HB (see Figure \ref{fig:parameters}). This is likely due to an intrinsic temperature change in \source over time. From Figure \ref{fig:parameters} we can see a clear variation in the temperatures over the past 10 years by $\sim$ 45\% and the focused observations happened when the system has a higher temperature.
To verify that the temperature change we see is not due to some analysis bias, we test three possible factors that could affect this result. 

The first one is the differences in the treatment of the hydrogen column density, $N_H$,  between the focused and SL spectra. The $N_H$ was tied between the branches in the focused data for each epoch but free to vary due to the increased statistical data quality of the spectra. However, the reduced signal-to-noise ratio of the SL data led to large variations in the inferred $N_H$, hence the parameter was fixed on the average of the focused datasets. Consequently, the lower $N_H$ used for the HB data in the focused spectra could lead to higher inferred disk temperatures as both components impact the shape of the lower-energy portion of the data. Fixing the $N_H$ value for the 2024 dataset (F04--F07) at the same value used for SL observations causes $kT_{bb}$ to decrease by about 0.03 keV. Hence, the difference in $N_H$ is insufficient to explain the observed temperature variation over the years. 

A second possibility could be the treatment of the background in the focused and SL observations. The aCXB and fCXB background components in SL spectra are the dominant background components below 10 keV. As a test of how these components impact the inferred temperatures, we removed the aCXB and fCXB component while keeping the other background components fixed. This results in a difference of less 0.3 keV in $kT_{bb}$ and $T_{in}$, which may explain some of the temperature variations between SL observations but it is not significant enough to explain the observed temperature difference between focused and SL data. 

A third possibility could be the calibration for SL observations. We check if SL and focused data will inherently provide different inferred values for disk temperature by looking at an intentional SL observation (ObsID 10402606002) that was conducted directly prior to a focused observation (ObsID 90401309029) of the BH X-ray binary (XRB) MAXI J1820+070, thus the data were obtained while the source was in the same soft spectral state. The data were both modeled with a simple continuum description ({\sc tbabs*simpl*diskbb}). The disk temperatures inferred are consistent within uncertainty (0.72--0.74 keV for SL and 0.72--0.73 keV for focused). Therefore, a difference in the calibration of the SL and focused data is not an issue. 

As a result, we are confident that the difference in the observed temperature between SL and focused observations are due to real temperature changes in the system over time. NS LMXBs do show temperature variations over time (e.g., GX340+0: \citealt{chatt_2024}, Cyg X-2: \citealt{ludlam22}, GX 13+1: \citealt{schnerr03, Kaddouh24}, the aforementioned plus Cir X-1 and XTE J1701-462: \citealt{fridriksson15}), which may arise from fluctuations in accretion rate or disk viscosity change \citep{kotze_2010,Durant_2009}, so this is not novel behavior. However, this highlights that SL observations can be very useful for studying the long-term properties of a system. This temperature change could be easily neglected if just considering the focused observations alone.

Although both focused and SL data show the same trend for $R_{BB}$ along the Z-track, the actual value in SL observations is larger than in focused observations. Again, this is likely due to the temperature change in the system. For most SL observations and some focused observations in FB, the spherical emission radius $R_{BB}$ value is too big to be inferred as the surface of the NS. However, the radius estimate could be affected by interstellar absorption, inclination angle, and whether a Comptonization component is included in the model. So it is expected that the estimation of radius is off from the actual radius of the NS. The inner disk radius $R_{in}$ from {\sc diskbb} shows a small increase from HB to NB and a significant increase in FB for focused observations. In SL observations, there is no clear trend for $R_{in}$ which is likely due to the large uncertainties on the normalization parameter.


Our results align with the findings of \cite{church2006}, demonstrating that the blackbody temperature ($kT_{bb}$) reaches its minimum value during the flaring branch (FB), while the blackbody radius ($R_{BB}$) achieves its maximum. This trend supports the earlier discussion of thermal properties varying along the Z-track and highlights the role of spectral modeling in capturing the accretion dynamics of \source.


\subsection{Reflection model}\label{sec:4p2}
In the focused observations, the blackbody and disk temperature decreases along the Z-track from HB to FB which is consistent with the continuum fitting results. Both focused observations in 2018 and 2024 give an inclination angle of $30^\circ$--$37^\circ$ which consistent with previous reports of \source being a low inclination system \citep{fender00,DA_2009,Cackett_2010,Miller_2016}. 
The inner disk radius \rin inferred from \relxillns are below 2.5\risco for all focused observations. This indicates that the system is not highly truncated, in agreement with the results from \cite{DA_2009}, \cite{Cackett_2010}, and \cite{chatt_2024}. Assuming a canonical NS mass of 1.4 $M_{\odot}$, the reflection fit gives an inner disk radius of $12.7$--$21.1$ km. If the truncation of the disk is due to the magnetic field, we can assume that the inner disk radius corresponds to the Alfv\'{e}n radius (to location where the energy density of the magnetic field balances that of the accreting material in the disk).  Using Eq. (1) from \cite{Cackett_2009}, assuming a distance of 11 kpc \citep{Fender_2000}, accretion efficiency of $\eta=0.2$, angular anisotropy $f_{ang}=1$, and conversion factor $kA=1$, an upper limit of the magnitude field strength can be estimated to be $B < 1.1 \times 10^9$ G at the magnetic poles. However, if the magnetosphere is not the responsible for the truncation, an extended BL from the NS surface could disrupt the accretion flow. Using Eq. (25) from \cite{Popham_2001}, using the unabsorbed flux from 0.5 -- 50 keV, the maximum radial extent of the BL ranges from $\sim 9.8 \ R_g-15.8\ Rg$ depending on the spectral state. Theses values are generally consistent with the truncation of the inner disk inferred from reflection modeling. A direct comparison of The maximum radial extent of the BL of each focused spectrum and \rin is given in \ref{tab:BL}.

\begin{figure}[t!]
\begin{center}
\includegraphics[width=0.48\textwidth, trim=1pt 1pt 0pt 0pt, clip]{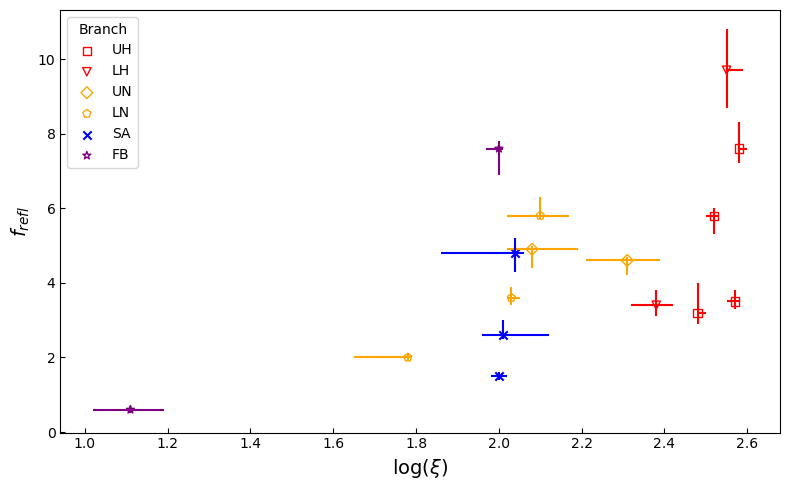} 
\caption{The relation between reflection fraction ($f_{refl}$) and ionization parameter ($\log(\xi)$) from the focused observations. The Spearman rank correlation factor is 0.55, indicating that there may be a weak correlation between higher reflection fractions and increased ionization.}
\label{ionfrac}
\end{center}
\end{figure}

The ionization parameter, $\log(\xi)$, is typically around $2$--$2.6$ which indicates that the system is moderately ionized. The iron abundance, $A_{Fe}$, is consistent with solar values with a few observations being high up to 4.31$\times$ solar. This parameter is known to be dependent on the disk density, $\log(n_e/\rm cm^{-3})$, which is fixed at maximum allowed value for the model \citep{García_2022}. The reflection fraction $f_{refl}$ spans a large range of values in the fit (0.5--10.8).  While this parameter provides insight into the system's reflected emission strength, it can fluctuate due to various factors and thus we do not directly infer any specific geometric constraints. The large reflection fraction could be due to the fact that the BL between the NS's surface and the accretion disk lies in the plane of the disk and is deep within the gravitational potential well of the NS, thus photons emitted from this region tend to be gravitationally focused onto the disk rather than escaping to infinity, similar to BH systems (see figure 2 in \cite{Dauser_2014} for when the corona is close to a rapidly spinning BH). This strong gravitational focusing enhances the amount of X-rays that are reflected off the disk, leading to a higher reflection fraction. In addition, the large variation of $f_{refl}$ could be due to the fact that the expected disk density $\log(n_e/\rm cm^{-3})$ for the system is higher than the maximum allowed value of $\log(n_e/\rm cm^{-3})=19$ in the model, thus other parameters may compensate to produce the same spectral features \citep{garcia18}. Furthermore, there is a weak relation between $\log(\xi)$ and $f_{refl}$ (see Figure~\ref{ionfrac}) with a spearman correlation factor of 0.55. 
We note that this is not a degeneracy of the model itself as shown in Figure \ref{fig:corner}. However, these parameters could be compensating for disk density. Another possible explanation is that when the reflection fraction is higher, there are more X-ray photons illuminating the disk, which increases the ionization parameter. As the ionization increases, the iron in the disk become more ionized producing weaker and broader lines, which makes them harder to detect so the model converges on higher reflection fraction to boost the contribution from the reflected emission.

In the SL observations, all the reflection parameters are poorly constrained. This is likely due to the lower signal-to-noise of SL data, especially when dividing the data into their respective branches. The total counts of SL observations are $\sim10\times$ lower than the total counts in focused observations. The highest total count in a single branch for SL observations is about 275k while the highest total count in a single branch for focused observations is about 3428k with most are above 1000k. Since SL observations have a lower count rate compared to focused observations, it is unsurprising that the reflection parameters are comparatively poorly constrained. However, we expect this to be significantly improved with intentional SL observations where the count rate is more than $10\times$ higher. With the significant increase in photon counts, we can not only model the reflection features more accurately, but also gain additional information at higher energies.\\

\section{Conclusion}
We present a comprehensive \nustar spectral analysis of 25 SL observations and 7 focused observations of the bright, persistently accreting NS LMXB \source to demonstrate that SL data provides a valuable complement to focused observations. The continuum model shows that the blackbody temperature remains almost the same from NB to HB, but has a significant drop in temperature while the blackbody emission radius significantly increases in the FB. The reflection model reaffirms that this is a low inclination system consistent with literature. In all branches, the system is moderately ionized, and the truncation is low with an estimated inner disk radius of 12.7--21.1~km. The SL data offer significant long-term temporal coverage starting from 2013, which enabled the study of variations in the thermal component of the system for over 10 years and contributed critical spectral insights which could be neglected without inclusion of SL data. Our spectral analysis showed consistency between the SL and focused data, particularly in the continuum model, although some differences were observed in the reflection model parameters, likely due to the lower quality of the SL data. These findings highlight the potential of SL observations to expand the scientific return from \nustar, particularly for sources where focused observations may be limited by telemetry constraints. Future work could focus on improving the constraints of the reflection model using intentional SL data with longer exposure times and higher photon counts, thereby improving the spectral fits and providing a deeper understanding of bright, accreting X-ray binary systems.\\

\noindent {\it Acknowledgments:} This work is supported by NASA under grant No.\ 80NSSC23K0498. Contributions by A.~Di Marco and F.~La Monaca are supported by the Istituto Nazionale di Astrofisica (INAF) and the Italian Space Agency (Agenzia Spaziale Italiana, ASI) through contract ASI-INAF-2022-19-HH.0.

\bibliography{reference} 
\bibliographystyle{aasjournal}
\FloatBarrier
\clearpage
\appendix

\renewcommand{\thesection}{\Roman{section}}

\section{Appendix}

\setcounter{table}{0}
\renewcommand{\thetable}{A\arabic{table}}

\FloatBarrier
\begin{longtable}[ht]{cccccccccc}
\caption{Information for all observations}
\label{tab:staycats_info}\\
\toprule
obs\# &  & ObsID   & Module & Area (arcsec) & Date & State &$C_{\rm Tot}$ ($10^3$) &$ T_{\rm exp}$ (ks) & Energy Range (keV)\\
\midrule
\endfirsthead

\multicolumn{9}{c}%
{{\bfseries \tablename\ \thetable{} -- continued from previous page}} \\
\toprule
Obs name & StrayID & OBSID   & Module & Area (cm$^2$) & Date & State &$C_{\rm Tot}$ ($10^3$) &$ T_{\rm exp}$ (ks)& Energy Range\\
\midrule
\endhead

\midrule
\multicolumn{8}{r}{{Continued on next page}} \\
\endfoot

\endlastfoot
F01&&30302030002 & A & 100 & 2018-08-12 & LN & 769.393 & 3.55 &3--28\\
&& & A & 100 & & SA & 2076.225 & 11.02&3--27 \\
&& & A & 100 & & FB & 663.404 & 2.89&3--18 \\
&& & B & 100 &  & LN & 512.896 & 2.47 &3--28\\
&& & B & 100 &  & SA & 2275.407 & 12.76 &3--27\\
&& & B & 100 &  & FB & 579.134 & 2.65 &3--18\\
F02&&30302030004 & A & 100 & 2018-08-15 & UN & 1807.626 & 6.67 &3--29\\
&& & A &100  &  & LN & 954.252 & 4.21&3--28 \\
 &&& A &  100&  & SA & 614.063 & 3.31&3--25 \\
 &&& A & 100 &  & FB & 259.554 & 1.15 &3--18\\
&& & B &100  && UN & 1693.510 & 6.68 &3--29\\
&& & B & 100 &  & LN & 888.359 & 4.12 &3--28\\
&& & B & 100 & & SA & 717.136 & 4.03 &3--25\\
&& & B & 100 & & FB & 198.966 & 0.93 &3--18\\
F03&&30302030006 & A & 100 & 2018-08-17 & UN & 3427.761 & 12.33&3--30\\
&& & A &  100& & LN & 380.190 & 1.65&3--25\\
 &&& A & 100 && SA & 182.391 & 0.90 &3--25\\
&& & B & 100 & & UN & 3171.744 & 12.15&3--30\\
&& & B &100  &  & LN & 370.797 & 1.65 &3--25\\
&& & B & 100 & & SA & 300.122 & 1.55&3--25\\

F04&&91002313002 & A &100 & 2024-03-27 & UH & 2743.064 & 12.46&3--30 \\
&& & B & 100 & & UH & 2604.428 & 12.81 &3--30\\
F05&&91002313004 & A & 100 & 2024-03-28 & UH & 3122.659 & 12.53 &3--30\\
&& & B &100  & & UH & 2896.329 & 12.93 &3--30\\
F06&&91002313006 & A &100  & 2024-03-31 & UH & 2409.635 & 9.27 &3--30\\
&& & A & 100 & & LN & 651.796 & 2.10 &3--30\\
&& & B &100  & & UH & 2750.941 & 11.15 &3--30\\
&& & B & 100 & & LN & 174.821 & 0.59&3--30\\
F07&&91002313008 & A &100  & 2024-04-01 & UH & 1269.084 & 5.47 &3--30\\
 &&& A & 100 && LN & 1699.523 & 5.48 &3--30\\
 &&& B & 100 & & UH & 1299.549 & 5.98 &3--30\\
 &&& B &100  &  & LN & 1536.652 & 5.34&3--30\\
\midrule
\midrule
obs\# & StrayID & ObsID   & Module & Area (cm$^2$) & Date & State &$C_{\rm Tot}$ ($10^3$) &$ T_{\rm exp}$ (ks)& Energe Range\\
\midrule
SL01 & StrayCatsI\_250 & 30001016002 & A & 2.58 & 2014-03-06 & HB & 21.095 & 7.34 & 3-20 \\
 &  &  &  &  &  & NB & 35.939 & 12.68 & 3-20 \\
SL02 & StrayCatsI\_254 & 30001012002 & A & 2.23 & 2013-03-23 & HB & 27.451 & 12.25 & 3-20 \\
 &  &  &  &  &  & NB & 7.666 & 3.44 & 3-20 \\
SL03 & StrayCatsI\_258 & 40014007001 & A & 2.15 & 2013-02-23 & NB & 19.296 & 9.25 & 3-20 \\
 &  &  &  &  &  & FB & 22.181 & 11.84 & 3-17 \\
SL04 & StrayCatsI\_260 & 40014024001 & A & 4.50 & 2014-03-11 & NB & 43.215 & 10.83 & 3-20 \\
 &  &  &  &  &  & FB & 56.519 & 15.83 & 3-14 \\
SL05 & StrayCatsI\_262 & 40014025001 & A & 4.30 & 2014-03-12 & NB & 30.419 & 9.01 & 3-20 \\
 &  &  &  &  &  & FB & 60.480 & 17.95 & 3-17 \\
SL06 & StrayCatsI\_264 & 30001017002 & A & 3.61 & 2014-05-12 & NB & 56.893 & 17.97 & 3-20 \\
 &  &  &  &  &  & FB & 84.838 & 29.17 & 3-17 \\
SL07 & StrayCatsI\_269 & 40014028002 & A & 4.43 & 2014-03-18 & HB & 49.073 & 10.57 & 3-20 \\
 &  &  &  &  &  & NB & 76.250 & 17.02 & 3-20 \\
SL08 & StrayCatsI\_273 & 30160003002 & A & 2.22 & 2015-05-31 & HB & 118.847 & 48.15 & 3-20 \\
 &  &  &  &  &  & NB & 63.922 & 24.79 & 3-20 \\
SL09 & StrayCatsI\_274 &  & B & 2.50 & 2015-05-31 & HB & 132.406 & 47.38 & 3-20 \\
 &  &  &  &  &  & NB & 76.968 & 25.94 & 3-20 \\
SL10 & StrayCatsI\_275 & 40014021002 & A & 4.58 & 2014-03-09 & HB & 15.892 & 3.27 & 3-20 \\
 &  &  &  &  &  & NB & 83.967 & 18.88 & 3-20 \\
 &  &  &  &  &  & FB & 22.137 & 5.79 & 3-17 \\
SL11 & StrayCatsI\_277 & 40014026001 & A & 5.47 & 2014-03-13 & HB & 14.528 & 2.74 & 3-20 \\
 &  &  &  &  &  & NB & 89.353 & 18.81 & 3-20 \\
 &  &  &  &  &  & FB & 34.620 & 7.81 & 3-17 \\
SL12 & StrayCatsI\_279 & 40014027001 & A & 5.25 & 2014-03-14 & HB & 14.980 & 2.96 & 3-20 \\
 &  &  &  &  &  & NB & 103.022 & 22.02 & 3-20 \\
 &  &  &  &  &  & FB & 17.367 & 4.37 & 3-17 \\
SL13 & StrayCatsI\_281 & 40014023001 & A & 4.97 & 2014-03-11 & NB & 60.779 & 14.12 & 3-20 \\
 &  &  &  &  &  & FB & 51.449 & 13.02 & 3-17 \\
SL14 & StrayCatsI\_283 & 30160001002 & A & 3.53 & 2015-06-11 & HB & 41.226 & 11.66 & 3-20 \\
 &  &  &  &  &  & NB & 52.099 & 15.84 & 3-20 \\
 &  &  &  &  &  & FB & 59.443 & 20.45 & 3-17 \\
SL15 & StrayCatsI\_285 &  & B & 3.88 & 2015-06-11 & HB & 36.076 & 9.27 & 3-20 \\
 &  &  &  &  &  & NB & 63.199 & 16.94 & 3-20 \\
 &  &  &  &  &  & FB & 69.698 & 21.70 & 3-17 \\
SL16 & StrayCatsI\_287 & 30001033002 & A & 6.23 & 2015-01-28 & HB & 275.047 & 41.28 & 3-20 \\
 &  &  &  &  &  & NB & 62.506 & 8.86 & 3-20 \\
SL17 & StrayCatsI\_290 & 40001022002 & A & 5.49 & 2014-03-07 & HB & 222.364 & 39.27 & 3-20 \\
 &  &  &  &  &  & NB & 266.840 & 48.84 & 3-20 \\
 &  &  &  &  &  & FB & 42.494 & 9.04 & 3-17 \\
SL18 & StrayCatsI\_292 & 30001008002 & B & 7.05 & 2014-06-26 & HB & 265.892 & 34.16 & 3-20 \\
 &  &  &  &  &  & NB & 110.066 & 13.87 & 3-20 \\
SL19 & StrayCatsI\_293 & 40014029001 & A & 7.38 & 2014-03-19 & HB & 160.875 & 23.37 & 3-20 \\
 &  &  &  &  &  & NB & 30.697 & 4.00 & 3-20 \\
SL20 & StrayCatsI\_349 & 30102054004 & A & 6.49 & 2015-08-31 & HB & 156.887 & 29.14 & 3-20 \\
SL21 & StrayCatsI\_351 & 30102054002 & A & 6.58 & 2015-08-21 & HB & 70.941 & 10.02 & 3-20 \\
 &  &  &  &  &  & NB & 77.911 & 11.42 & 3-18 \\
SL22 & StrayCatsI\_355 & 30102054006 & A & 6.47 & 2015-09-12 & NB & 39.572 & 7.30 & 3-20 \\
 &  &  &  &  &  & FB & 67.615 & 13.00 & 3-17 \\
SL23 & StrayCatsII\_29 & 90601324002 & A & 3.70 & 2020-07-31 & NB & 31.376 & 10.30 & 3-20 \\
 &  &  &  &  &  & FB & 39.112 & 13.42 & 3-17 \\
SL24 & StrayCatsIII\_41 & 90901311002 & A & 2.41 & 2023-03-19 & HB & 33.384 & 12.15 & 3-20 \\
 &  &  &  &  &  & NB & 24.759 & 8.98 & 3-20 \\
SL25 & StrayCatsIII\_43 &  & B & 2.47 & 2023-03-19 & HB & 31.168 & 10.71 & 3-20 \\
 &  &  &  &  &  & NB & 29.877 & 10.32 & 3-20 \\
\bottomrule
\end{longtable}
\renewcommand{\arraystretch}{1.2}
\setlength{\LTcapwidth}{\textwidth}
\begin{center}
\begin{longtable}{ccccccccc}

\caption[\textbf{Continuum fit result for all stray light observations}]{
    Continuum fit result for all stray light observations. 
    }

\label{continuum_full} \\
\toprule
& & \multicolumn{3}{c}{\sc bbody} & \multicolumn{3}{c}{\sc diskbb} &  \\
\cmidrule(lr){3-5}
\cmidrule(lr){6-8}
obs\# & Branch & $kT_{bb}\ \rm (keV)$ & norm($10^{-2}$) & $R_{BB}\ \rm (km)$ & $T_{in}\ \rm (keV)$ & norm & $R_{in}$ (km) & C/dof \\
\midrule
\endfirsthead

\caption[]{Continuum fit result for all stray light observations (continued)} \\
\toprule
& & \multicolumn{3}{c}{\sc bbody} & \multicolumn{3}{c}{\sc diskbb} &  \\
\cmidrule(lr){3-5}
\cmidrule(lr){6-8}
obs\# & Branch & $kT_{bb}\ \rm (keV)$ & norm ($10^{-2}$) & $R_{BB}\ \rm (km)$ & $T_{in}\ \rm (keV)$ & norm & $R_{in}$ (km) & C/dof \\
\midrule
\endhead

\bottomrule
\multicolumn{9}{r}{\textit{Continued on next page...}} \\
\endfoot

\bottomrule
\endlastfoot
SL01 & HB & $2.19^{+0.11}_{-0.09}$ & $11.1^{+1.2}_{-1.3}$ & $114^{+34}_{-33}$ & $1.21^{+0.10}_{-0.09}$ & $561^{+226}_{-155}$ & $49^{+17}_{-15}$ & $56/56$ \\
& NB & $2.19^{+0.2}_{-0.1}$ & $8.7^{+1.6}_{-1.7}$ & $102^{+33}_{-31}$ & $1.37^{+0.11}_{-0.11}$ & $332^{+124}_{-86}$ & $37^{+12}_{-12}$ & $86/60$\\
SL02 & HB & $2.42^{+0.11}_{-0.09}$ & $9.3^{+0.8}_{-0.9}$ & $86^{+25}_{-24}$ & $1.27^{+0.08}_{-0.07}$ & $397^{+119}_{-89}$ & $41^{+12}_{-12}$ & $58/59$\\
& NB & $2.5^{+0.4}_{-0.3}$ & $7.7^{+1.8}_{-1.9}$ & $75^{+30}_{-26}$ & $1.38^{+0.2}_{-0.2}$ & $281^{+194}_{-106}$ & $34^{+15}_{-12}$ & $48/51$\\
SL03 & NB & $2.0^{+0.2}_{-0.1}$ & $7.2^{+1.5}_{-1.5}$ & $112^{+37}_{-35}$ & $1.2^{+0.1}_{-0.1}$ & $580^{+247}_{-163}$ & $49^{+17}_{-15}$ & $78/55$\\
& FB & $1.7^{+0.3}_{-0.2}$ & $7.5^{+2.7}_{-3.3}$ & $157^{+70}_{-60}$ & $1.12^{+0.2}_{-0.2}$ & $710^{+691}_{-328}$ & $54^{+30}_{-19}$ & $85/50$\\
SL04 & NB & $1.79^{+0.08}_{-0.06}$ & $8.5^{+1.1}_{-1.2}$ & $149^{+44}_{-43}$ & $1.08^{+0.08}_{-0.08}$ & $739^{+296}_{-201}$ & $56^{+19}_{-17}$ & $59/60$\\
& FB & $1.6^{+1.0}_{-1.2}$ & $9.0^{+14}_{-0.4}$ & $190^{+279}_{-285}$ & $0.8^{+0.1}_{-0.1}$ & $982^{+880}_{-911}$ & $64^{+35}_{-35}$ & $36/43$\\
SL05 & NB & $1.52^{+0.05}_{-0.04}$ & $10.3^{+0.9}_{-0.9}$ & $227^{+64}_{-64}$ & $0.83^{+0.08}_{-0.07}$ & $2230^{+1360}_{-808}$ & $97^{+39}_{-32}$ & $70/58$\\
& FB & $1.37^{+0.02}_{-0.02}$ & $12.7^{+0.5}_{-0.6}$ & $311^{+86}_{-86}$ & $0.66^{+0.06}_{-0.05}$ & $7348^{+4819}_{-2802}$ & $175^{+75}_{-58}$ & $49/52$\\
SL06 & NB & $1.67^{+0.05}_{-0.04}$ & $9.8^{+0.9}_{-0.9}$ & $185^{+52}_{-52}$ & $0.96^{+0.07}_{-0.07}$ & $1259^{+524}_{-355}$ & $73^{+26}_{-22}$ & $68/61$\\
& FB & $1.34^{+0.02}_{-0.02}$ & $13.4^{+0.5}_{-0.5}$ & $335^{+92}_{-92}$ & $0.64^{+0.05}_{-0.05}$ & $9137^{+5826}_{-3409}$ & $196^{+82}_{-65}$ & $55/53$\\
SL07 & HB & $2.11^{+0.07}_{-0.06}$ & $10.3^{+0.9}_{-0.9}$ & $119^{+34}_{-33}$ & $1.19^{+0.07}_{-0.07}$ & $536^{+160}_{-119}$ & $47^{+15}_{-13}$ & $99/61$\\
& NB & $2.11^{+0.09}_{-0.07}$ & $8.1^{+0.9}_{-1.0}$ & $105^{+30}_{-30}$ & $1.29^{+0.07}_{-0.07}$ & $381^{+93}_{-72}$ & $39^{+12}_{-12}$ & $133/63$ \\
SL08 & HB & $2.28^{+0.05}_{-0.04}$ & $10.9^{+0.6}_{-0.6}$ & $104^{+29}_{-29}$ & $1.27^{+0.05}_{-0.04}$ & $410^{+67}_{-56}$ & $41^{+12}_{-12}$ & $152/68$ \\
& NB & $2.31^{+0.13}_{-0.10}$ & $9.0^{+1.3}_{-1.3}$ & $92^{+28}_{-27}$ & $1.42^{+0.09}_{-0.09}$ & $278^{+80}_{-59}$ & $34^{+10}_{-10}$ & $168/65$\\

SL09 & HB & $2.22^{+0.04}_{-0.04}$ & $11.6^{+0.5}_{-0.5}$ & $114^{+31}_{-31}$ & $1.21^{+0.04}_{-0.04}$ & $477^{+80}_{-68}$ & $44^{+12}_{-12}$ & $137/68$\\
& NB & $2.27^{+0.10}_{-0.08}$ & $9.85^{+1.1}_{-1.2}$ & $101^{+29}_{-29}$ & $1.38^{+0.09}_{-0.08}$ & $308^{+85}_{-65}$ & $36^{+12}_{-10}$ & $208/63$\\
SL10 & HB & $1.99^{+0.09}_{-0.08}$ & $11.41^{+1.2}_{-1.4}$ & $140^{+41}_{-41}$ & $1.07^{+0.11}_{-0.10}$ & $887^{+498}_{-305}$ & $61^{+24}_{-20}$ & $62/55$ \\
& NB & $1.88^{+0.06}_{-0.05}$ & $9.09^{+0.8}_{-0.9}$ & $140^{+40}_{-39}$ & $1.13^{+0.06}_{-0.06}$ & $687^{+173}_{-133}$ & $54^{+17}_{-15}$ & $116/64$\\
& FB & $1.42^{+0.05}_{-0.04}$ & $12.36^{+1.1}_{-1.5}$ & $288^{+82}_{-82}$ & $0.76^{+0.12}_{-0.10}$ & $3474^{+4341}_{-1841}$ & $121^{+82}_{-46}$ & $56/47$ \\
SL11 & HB & $2.06^{+0.18}_{-0.13}$ & $8.54^{+1.8}_{-2.0}$ & $114^{+39}_{-37}$ & $1.26^{+0.16}_{-0.15}$ & $389^{+263}_{-147}$ & $41^{+17}_{-14}$ & $88/55$\\
& NB & $1.78^{+0.05}_{-0.04}$ & $8.87^{+0.7}_{-0.8}$ & $155^{+44}_{-43}$ & $1.05^{+0.06}_{-0.05}$ & $794^{+211}_{-161}$ & $58^{+17}_{-17}$ & $104/64$ \\
& FB & $1.38^{+0.03}_{-0.03}$ & $12.98^{+0.8}_{-0.9}$ & $311^{+86}_{-87}$ & $0.73^{+0.10}_{-0.08}$ & $3393^{+3423}_{-1613}$ & $119^{+68}_{-42}$ & $65/50$\\
SL12 & HB & $2.03^{+0.14}_{-0.11}$ & $9.1^{+1.5}_{-1.6}$ & $120^{+38}_{-37}$ & $1.17^{+0.13}_{-0.12}$ & $528^{+330}_{-191}$ & $48^{+20}_{-15}$ & $57/54$\\
& NB & $1.82^{+0.04}_{-0.04}$ & $9.03^{+0.6}_{-0.7}$ & $149^{+42}_{-42}$ & $1.06^{+0.05}_{-0.05}$ & $780^{+186}_{-146}$ & $58^{+17}_{-17}$ & $101/65$\\
& FB & $1.44^{+0.07}_{-0.05}$ & $10.08^{+1.2}_{-1.6}$ & $250^{+74}_{-73}$ & $0.8^{+0.13}_{-0.11}$ & $2535^{+3061}_{-1346}$ & $104^{+68}_{-39}$ & $51/45$\\
SL13 & NB & $1.81^{+0.09}_{-0.07}$ & $7.15^{+1.1}_{-1.2}$ & $135^{+40}_{-40}$ & $1.15^{+0.08}_{-0.08}$ & $572^{+193}_{-137}$ & $49^{+15}_{-15}$ & $86/61$\\
& FB & $1.41^{+0.04}_{-0.03}$ & $11.21^{+0.9}_{-1.0}$ & $275^{+77}_{-77}$ & $0.81^{+0.09}_{-0.08}$ & $2368^{+1711}_{-951}$ & $100^{+46}_{-34}$ & $33/51$\\
SL14 & HB & $2.2^{+0.07}_{-0.06}$ & $10.74^{+0.7}_{-0.8}$ & $111^{+31}_{-31}$ & $1.16^{+0.07}_{-0.07}$ & $507^{+159}_{-118}$ & $46^{+14}_{-14}$ & $90/65$\\
& NB & $1.8^{+0.08}_{-0.06}$ & $8.81^{+1.1}_{-1.2}$ & $150^{+44}_{-43}$ & $1.1^{+0.09}_{-0.08}$ & $699^{+291}_{-195}$ & $54^{+19}_{-17}$ & $90/64$\\
& FB & $1.37^{+0.03}_{-0.02}$ & $12.67^{+0.6}_{-0.7}$ & $311^{+86}_{-86}$ & $0.69^{+0.07}_{-0.06}$ & $5839^{+4265}_{-2363}$ & $156^{+72}_{-53}$ & $44/54$\\
SL15 & HB & $2.19^{+0.08}_{-0.07}$ & $10.4^{+0.8}_{-0.9}$ & $110^{+31}_{-31}$ & $1.19^{+0.09}_{-0.08}$ & $429^{+159}_{-113}$ & $43^{+14}_{-14}$ & $85/61$\\
& NB & $1.91^{+0.07}_{-0.06}$ & $8.66^{+0.9}_{-0.9}$ & $132^{+38}_{-38}$ & $1.14^{+0.07}_{-0.07}$ & $588^{+178}_{-132}$ & $49^{+15}_{-15}$ & $106/62$\\
& FB & $1.39^{+0.02}_{-0.02}$ & $12.45^{+0.6}_{-0.6}$ & $301^{+83}_{-83}$ & $0.7^{+0.06}_{-0.05}$ & $5351^{+3380}_{-2001}$ & $150^{+63}_{-49}$ & $62/53$\\
SL16 & HB & $2.17^{+0.03}_{-0.03}$ & $11.2^{+0.4}_{-0.4}$ & $117^{+32}_{-32}$ & $1.22^{+0.03}_{-0.03}$ & $453^{+51}_{-45}$ & $44^{+12}_{-12}$ & $201/70$\\
& NB & $2.2^{+0.12}_{-0.09}$ & $9.65^{+1.4}_{-1.5}$ & $106^{+32}_{-31}$ & $1.38^{+0.11}_{-0.10}$ & $294^{+97}_{-71}$ & $36^{+12}_{-10}$ & $138/63$\\
SL17 & HB & $2.17^{+0.03}_{-0.03}$ & $10.43^{+0.4}_{-0.4}$ & $113^{+31}_{-31}$ & $1.21^{+0.03}_{-0.03}$ & $458^{+58}_{-51}$ & $44^{+12}_{-12}$ & $150/69$\\
& NB & $1.86^{+0.03}_{-0.03}$ & $9.74^{+0.5}_{-0.5}$ & $148^{+41}_{-41}$ & $1.12^{+0.04}_{-0.03}$ & $682^{+99}_{-85}$ & $53^{+15}_{-15}$ & $212/69$ \\
& FB & $1.43^{+0.03}_{-0.03}$ & $12.69^{+0.7}_{-0.8}$ & $286^{+79}_{-80}$ & $0.74^{+0.07}_{-0.06}$ & $4288^{+2813}_{-1642}$ & $134^{+58}_{-45}$ & $66/52$\\

SL18 & HB & $2.11^{+0.03}_{-0.03}$ & $11.42^{+0.4}_{-0.4}$ & $125^{+34}_{-34}$ & $1.18^{+0.03}_{-0.03}$ & $513^{+62}_{-55}$ & $46^{+14}_{-14}$ & $178/73$ \\
& NB & $1.98^{+0.05}_{-0.05}$ & $11.12^{+0.8}_{-0.9}$ & $140^{+39}_{-39}$ & $1.18^{+0.07}_{-0.06}$ & $548^{+139}_{-108}$ & $48^{+14}_{-14}$ & $131/66$\\
SL19 & HB & $2.23^{+0.04}_{-0.04}$ & $9.15^{+0.37}_{-0.39}$ & $100^{+28}_{-28}$ & $1.22^{+0.04}_{-0.04}$ & $366^{+60}_{-51}$ & $39^{+12}_{-10}$ & $163/69$\\
& NB &$1.92^{+0.09}_{-0.07}$ & $10.7^{+1.2}_{-1.3}$ & $145^{+43}_{-42}$ & $1.10^{+0.10}_{-0.09}$ & $696^{+335}_{-219}$ & $54^{+20}_{-17}$ & $73/60$ \\
SL20 & HB & $2.25^{+0.04}_{-0.03}$ & $8.91^{+0.28}_{-0.30}$ & $97^{+27}_{-27}$ & $1.16^{+0.04}_{-0.04}$ & $376^{+64}_{-54}$ & $39^{+12}_{-10}$ & $162/69$\\
SL21 & HB &$2.04^{+0.05}_{-0.04}$ & $11.8^{+0.6}_{-0.7}$ & $135^{+38}_{-38}$ & $1.12^{+0.06}_{-0.06}$ & $652^{+173}_{-134}$ & $53^{+15}_{-15}$ & $76/63$ \\
& NB & $1.87^{+0.06}_{-0.05}$ & $10.5^{+0.9}_{-1.0}$ & $153^{+43}_{-43}$ & $1.10^{+0.07}_{-0.07}$ & $738^{+241}_{-176}$ & $56.1^{+18}_{-17}$ & $92/57$ \\
SL22 & NB & $1.61^{+0.08}_{-0.06}$ & $8.03^{+1.2}_{-1.3}$ & $180^{+54}_{-53}$ & $1.00^{+0.09}_{-0.09}$ & $1032^{+543}_{-333}$ & $66^{+26}_{-20}$ & $59/59$\\
& FB & $1.31^{+0.02}_{-0.02}$ & $13.0^{+0.6}_{-0.7}$ & $343^{+94}_{-95}$ & $0.65^{+0.06}_{-0.06}$ & $7408^{+5848}_{-3096}$ & $176^{+85}_{-61}$ & $85/52$\\
SL23 & NB &$1.81^{+0.10}_{-0.08}$ & $7.01^{+1.1}_{-1.2}$ & $133^{+41}_{-40}$ & $1.13^{+0.09}_{-0.08}$ & $604^{+240}_{-161}$ & $51^{+17}_{-15}$ & $83/59$ \\
& FB &$1.42^{+0.04}_{-0.03}$ & $11.7^{+0.7}_{-0.9}$ & $278^{+78}_{-77}$ & $0.75^{+0.08}_{-0.07}$ & $3627^{+2721}_{-1488}$ & $124^{+58}_{-42}$ & $64/51$ \\
SL24 & HB & $2.19^{+0.09}_{-0.07}$ & $11.2^{+1.0}_{-1.1}$ & $115^{+33}_{-33}$ & $1.23^{+0.08}_{-0.08}$ & $476^{+154}_{-113}$ & $44^{+14}_{-14}$ & $92/64$\\
& NB & $2.26^{+0.16}_{-0.12}$ & $9.07^{+1.5}_{-1.6}$ & $97^{+31}_{-30}$ & $1.35^{+0.11}_{-0.11}$ & $352^{+138}_{-94}$ & $39^{+14}_{-12}$ & $103/64$\\
SL25 & HB & $2.19^{+0.08}_{-0.07}$ & $11.6^{+0.9}_{-1.0}$ & $116^{+33}_{-33}$ & $1.20^{+0.08}_{-0.07}$ & $546^{+177}_{-130}$ & $48^{+15}_{-14}$ & $77/60$\\
& NB & $2.32^{+0.18}_{-0.14}$ & $8.39^{+1.5}_{-1.6}$ & $89^{+29}_{-28}$ & $1.40^{+0.11}_{-0.11}$ & $299^{+112}_{-76}$ & $36^{+12}_{-10}$ & $100/60$\\
\end{longtable}
\end{center}
\vspace{-20pt}
Note. -- The column density $N_H\ (10^{22}\ \mathrm{cm}^{-2})$ is fixed at 6.39, which is the average value found in Table~\ref{tab:reflection_focused}. $R_{BB}$ and $R_{in}$ are calculated using the same values for $f_{cor}$, distance, and inclination as Table~\ref{tab:Continuum_focused}.


\setlength{\tabcolsep}{2pt}
\begin{longtable}{p{0.5cm}ccccccccccp{0.05cm}}
\caption{Reflection fit for all SL observations} \label{tab:reflection_all} \\

\toprule
\multicolumn{2}{c}{} & \multicolumn{6}{c}{\sc relxillNS} & \multicolumn{2}{c}{{\sc diskbb}} &  \\
\cmidrule(lr){3-8} \cmidrule(lr){9-10}
obs\# & Branch  & $R_{in}$ (\risco) & $kT_{bb}$ (keV) & $\log(\xi)$ & $A_{Fe}$ & $f_{refl}$&$\rm norm\ (10^{-3})$&$T_{in}$ (keV)&norm& $C/dof$ \\
\midrule
\endfirsthead

\multicolumn{11}{c}{{\tablename\ \thetable{} - continued from previous page}} \\
\toprule
\multicolumn{2}{c}{} & \multicolumn{6}{c}{\sc relxillNS} & \multicolumn{2}{c}{\sc diskbb} &  \\
\cmidrule(lr){3-8} \cmidrule(lr){9-10}
obs\# & Branch  & $R_{in}$ (\risco) & $kT_{bb}$ (keV) & $\log(\xi)$ & $A_{Fe}$ & $f_{refl}$&$\rm norm\ (10^{-3})$&$T_{in}$ (keV)&norm& $C/dof$ \\
\midrule
\endhead

\bottomrule
\multicolumn{11}{r}{{Continued on next page}} \\
\endfoot

\bottomrule
\endlastfoot
SL01 & HB &${9.65}_{-8.30}^{+0.08}$ & ${2.38}^{+0.02}_{-0.26}$ & ${4.2}^{+0.4}_{-2.8}$ & ${6.8}^{+2.8}_{-5.7}$ & ${1.82}^{+0.15}_{-1.68}$ & ${4.3}^{+7.2}_{-0.1}$ & ${1.08}^{+0.19}_{-0.08}$ & ${834}^{+346}_{-414}$ & $47/52$\\
& NB &${9.66}_{-7.28}^{+0.06}$ & ${2.16}^{+1.07}_{-0.07}$ & ${3.0}^{+1.5}_{-0.6}$ & ${9.7}^{+0.1}_{-9.1}$ & ${0.23}^{+2.95}_{-0.06}$ & ${9.2}^{+0.7}_{-8.0}$ & ${1.27}^{+0.49}_{-0.17}$ & ${417}^{+315}_{-291}$ & $59/56$\\
SL02 & HB &${6.46}_{-5.06}^{+3.07}$ & ${2.53}^{+0.26}_{-0.09}$ & ${4.6}^{+0.1}_{-0.9}$ & ${9.7}^{+0.2}_{-7.8}$ & ${0.76}^{+4.65}_{-0.32}$ & ${6.0}^{+1.7}_{-4.5}$ & ${1.14}^{+0.12}_{-0.09}$ & ${575}^{+240}_{-205}$ & $45/55$\\
& NB &${7.47}_{-6.01}^{+2.06}$ & ${2.43}^{+0.46}_{-0.14}$ & ${3.5}^{+1.1}_{-0.8}$ & ${4.9}^{+4.6}_{-3.9}$ & ${0.46}^{+4.74}_{-0.31}$ & ${7.0}^{+1.1}_{-5.6}$ & ${1.21}^{+0.27}_{-0.23}$ & ${443}^{+696}_{-243}$ & $44/47$\\
SL03 & NB &${7.6}_{-6.2}^{+1.9}$ & ${2.13}^{+0.12}_{-0.20}$ & ${4.1}^{+0.6}_{-2.7}$ & ${1.2}^{+8.2}_{-0.3}$ & ${1.30}^{+0.03}_{-1.27}$ & ${3.2}^{+4.5}_{-0.1}$& ${1.19}^{+0.18}_{-0.13}$ & ${547}^{+374}_{-223}$ & $76/51$\\
& FB&${8.75}_{-7.08}^{+0.84}$ & ${1.76}^{+0.13}_{-0.15}$ & ${2.8}^{+1.7}_{-0.9}$ & ${5.4}^{+4.0}_{-4.4}$ & ${0.53}^{+2.09}_{-0.17}$ & ${6.2}^{+2.1}_{-3.6}$ & ${1.08}^{+0.06}_{-0.25}$ & ${796}^{+1703}_{-204}$ & $55/44$ \\
SL04 & NB&${1.05}_{-0.25}^{+8.16}$ & ${1.88}^{+0.11}_{-0.08}$ & ${3.7}^{+1.0}_{-1.4}$ & ${1.8}^{+7.4}_{-1.1}$ & ${1.50}^{+0.33}_{-1.28}$ & ${4.5}^{+3.4}_{-1.4}$ & ${0.90}^{+0.24}_{-0.02}$ & ${1438}^{+58}_{-912}$ & $43/56$ \\
& FB&${1.9}_{-0.5}^{+7.7}$ & ${1.41}^{+0.06}_{-0.03}$ & ${2.0}^{+2.6}_{-0.9}$ & ${5.1}^{+4.3}_{-4.3}$ & ${0.01}^{+0.32}_{-0.01}$ & ${11.7}^{+0.6}_{-2.9}$ & ${0.81}^{+0.08}_{-0.09}$ & ${2205}^{+1780}_{-773}$ & $36/39$ \\
SL05 & NB&${9.9}_{-8.6}^{+0.4}$ & ${1.57}^{+0.04}_{-0.07}$ & ${2.4}^{+1.7}_{-1.4}$ & ${8.3}^{+1.2}_{-7.8}$ & ${0.26}^{+18.44}_{-0.13}$ & ${9.1}^{+1.1}_{-8.0}$ & ${0.90}^{+0.16}_{-0.11}$ & ${1436}^{+1224}_{-630}$ & $62/54$ \\
& FB& ${2.48}_{-1.07}^{+6.95}$ & ${1.40}^{+0.04}_{-0.04}$ & ${2.5}^{+1.9}_{-1.2}$ & ${7.7}^{+1.7}_{-7.1}$ & ${0.17}^{+1.43}_{-0.13}$ & ${11.8}^{+1.1}_{-5.1}$ & ${0.74}^{+0.12}_{-0.08}$ & ${3558}^{+3541}_{-1882}$ & $45/48$\\
SL06 & NB&${1.5}_{-0.2}^{+7.8}$ & ${1.88}^{+0.12}_{-0.18}$ & ${3.7}^{+0.9}_{-0.5}$ & ${0.6}^{+2.4}_{-0.1}$ & ${10}^{+5}_{-6}$ & ${0.7}^{+2.2}_{-0.3}$ & ${0.84}^{+0.15}_{-0.13}$ & ${2180}^{+691}_{-1141}$ & $50/57$ \\
& FB&${1.42}_{-0.10}^{+8.02}$ & ${1.35}^{+0.02}_{-0.09}$ & ${1.5}^{+2.9}_{-0.4}$ & ${1.9}^{+7.5}_{-1.0}$ & ${0.14}^{+0.91}_{-0.13}$ & ${13.3}^{+6.6}_{-1.1}$ & ${0.67}^{+0.13}_{-0.06}$ & ${6119}^{+5014}_{-3339}$ & $56/49$ \\
SL07 & HB &${9.78}_{-8.61}^{+0.03}$ & ${2.10}^{+0.25}_{-0.04}$ & ${2.1}^{+1.8}_{-0.4}$ & ${1.5}^{+6.7}_{-0.9}$ & ${0.28}^{+2.57}_{-0.06}$ & ${10.0}^{+0.6}_{-7.2}$ & ${1.09}^{+0.10}_{-0.08}$ & ${579}^{+397}_{-104}$ & $66/57$\\
& NB& ${5.8}_{-3.3}^{+4.0}$ & ${2.20}^{+0.35}_{-0.11}$ & ${3.8}^{+0.7}_{-2.0}$ & ${3.4}^{+6.1}_{-2.8}$ & ${1.60}^{+1.95}_{-1.42}$ & ${3.9}^{+4.3}_{-2.5}$ & ${1.11}^{+0.45}_{-0.05}$ & ${639}^{+147}_{-461}$ & $97/59$\\
SL08 & HB& ${1.14}_{-0.04}^{+5.98}$ & ${2.43}^{+0.05}_{-0.14}$ & ${4.1}^{+0.5}_{-1.4}$ & ${8.6}^{+1.1}_{-7.9}$ & ${1.47}^{+0.65}_{-1.43}$ & ${5.5}^{+5.2}_{-1.3}$ & ${1.04}^{+0.30}_{-0.03}$ & ${869}^{+113}_{-567}$ & $88/64$\\
& NB& ${4.4}_{-3.2}^{+5.2}$ & ${2.43}^{+0.16}_{-0.21}$ & ${3.8}^{+0.8}_{-0.8}$ & ${3.6}^{+1.1}_{-3.0}$ & ${10}^{+3}_{-10}$ & ${1.0}^{+5.9}_{-0.2}$ & ${1.00}^{+0.38}_{-0.04}$ & ${1131}^{+233}_{-851}$ & $91/59$\\

SL09 & HB&${7.9}_{-6.8}^{+1.6}$ & ${2.25}^{+0.27}_{-0.03}$ & ${3.3}^{+0.5}_{-2.3}$ & ${9.5}^{+0.0}_{-9.0}$ & ${0.18}^{+4.95}_{-0.02}$ & ${11.4}^{+0.3}_{-9.1}$ & ${1.12}^{+0.40}_{-0.07}$ & ${617}^{+205}_{-429}$ & $61/64$ \\
& NB &${5.0}_{-3.8}^{+4.4}$ & ${2.29}^{+0.24}_{-0.04}$ & ${4.4}^{+0.2}_{-0.5}$ & ${9.3}^{+0.5}_{-7.7}$ & ${1.63}^{+2.42}_{-0.34}$ & ${5.1}^{+0.8}_{-2.5}$ & ${1.01}^{+0.34}_{-0.08}$ & ${1058}^{+489}_{-754}$ & $128/59$\\
SL10 & HB&${7.8}_{-6.5}^{+1.7}$ & ${2.03}^{+0.12}_{-0.11}$ & ${1.9}^{+2.7}_{-0.7}$ & ${1.2}^{+8.2}_{-0.2}$ & ${0.09}^{+0.63}_{-0.07}$ & ${11.0}^{+1.3}_{-3.8}$ & ${1.10}^{+0.11}_{-0.17}$ & ${672}^{+980}_{-188}$ & $61/51$ \\
& NB&${5.7}_{-4.2}^{+3.9}$ & ${2.04}^{+0.04}_{-0.14}$ & ${3.8}^{+0.9}_{-1.5}$ & ${3.1}^{+6.4}_{-1.8}$ & ${1.57}^{+0.50}_{-1.27}$ & ${4.0}^{+4.4}_{-0.7}$ & ${1.01}^{+0.07}_{-0.06}$ & ${1046}^{+344}_{-342}$ & $58/60$ \\
& FB &${9.0}_{-7.7}^{+0.6}$ & ${1.50}^{+0.07}_{-0.09}$ & ${4.5}^{+0.1}_{-3.2}$ & ${7.9}^{+1.3}_{-7.2}$ & ${0.27}^{+0.89}_{-0.22}$ & ${9.2}^{+3.3}_{-3.8}$ & ${0.80}^{+0.17}_{-0.11}$ & ${2511}^{+3156}_{-1498}$ & $53/43$\\
SL11 & HB& ${8.53}_{-7.03}^{+1.09}$ & ${2.06}^{+0.17}_{-0.10}$ & ${3.0}^{+1.5}_{-0.9}$ & ${9.6}^{+0.1}_{-8.3}$ & ${0.25}^{+2.60}_{-0.04}$ & ${8.8}^{+0.7}_{-5.9}$ & ${1.14}^{+0.12}_{-0.23}$ & ${558}^{+939}_{-210}$ & $77/51$\\
& NB& ${1.4}_{-0.1}^{+6.8}$ & ${1.81}^{+0.03}_{-0.09}$ & ${2.3}^{+0.7}_{-1.3}$ & ${1.4}^{+4.4}_{-0.9}$ & ${0.40}^{+16.45}_{-0.22}$ & ${7.6}^{+1.1}_{-6.6}$ & ${1.04}^{+0.19}_{-0.07}$ & ${664}^{+297}_{-235}$ & $69/60$\\
& FB& ${5.0}_{-3.6}^{+4.5}$ & ${1.39}^{+0.05}_{-0.03}$ & ${1.4}^{+3.1}_{-0.3}$ & ${3.1}^{+6.0}_{-2.5}$ & ${0.01}^{+0.81}_{-0.01}$ & ${13.3}^{+0.7}_{-5.1}$ & ${0.77}^{+0.20}_{-0.14}$ & ${2112}^{+4815}_{-1415}$ & $66/46$\\
SL12 & HB& ${9.4}_{-7.8}^{+0.4}$ & ${2.04}^{+0.29}_{-0.29}$ & ${2.7}^{+1.9}_{-0.5}$ & ${5.3}^{+3.9}_{-4.5}$ & ${0.28}^{+2.90}_{-0.04}$ & ${9.0}^{+5.3}_{-6.7}$ & ${1.08}^{+0.21}_{-0.16}$ & ${724}^{+719}_{-324}$ & $47/50$\\
& NB &${2.0}_{-0.7}^{+7.2}$ & ${1.93}^{+0.07}_{-0.19}$ & ${4.6}^{+0.0}_{-2.6}$ & ${9.6}^{+0.0}_{-8.9}$ & ${0.75}^{+0.94}_{-0.55}$ & ${5.8}^{+14.4}_{-2.1}$ & ${0.95}^{+0.12}_{-0.03}$ & ${1202}^{+173}_{-683}$ & $53/61$\\
& FB&${9.7}_{-7.6}^{+0.1}$ & ${1.53}^{+0.15}_{-0.10}$ & ${2.6}^{+1.9}_{-1.0}$ & ${6.0}^{+3.4}_{-5.0}$ & ${0.37}^{+1.19}_{-0.23}$ & ${0.01}^{+0.01}_{-0.01}$ & ${0.90}^{+0.10}_{-0.18}$ & ${1405}^{+2868}_{-556}$ & $43/41$ \\
SL13 & NB&${7.1}_{-5.7}^{+2.4}$ & ${2.04}^{+0.09}_{-0.07}$ & ${3.7}^{+1.0}_{-0.1}$ & ${0.6}^{+3.1}_{-0.0}$ & ${15}^{+4}_{-8}$ & ${0.5}^{+0.4}_{-0.1}$ & ${1.08}^{+0.09}_{-0.09}$ & ${695}^{+332}_{-210}$ & $69/57$ \\
& FB& ${2.9}_{-1.5}^{+6.6}$ & ${1.44}^{+0.08}_{-0.05}$ & ${2.0}^{+2.6}_{-0.9}$ & ${0.8}^{+8.4}_{-0.2}$ & ${1.04}^{+0.29}_{-1.01}$ & ${7.0}^{+4.6}_{-0.6}$ & ${0.96}^{+0.01}_{-0.24}$ & ${964}^{+2727}_{-38}$ & $31/47$\\
SL14 & HB& ${1.17}_{-0.07}^{+7.94}$ & ${2.17}^{+0.35}_{-0.01}$ & ${1.9}^{+1.7}_{-0.9}$ & ${1.4}^{+6.9}_{-0.9}$ & ${0.41}^{+9.24}_{-0.19}$ & ${10.2}^{+0.4}_{-8.9}$ & ${0.95}^{+0.61}_{-0.04}$ & ${746}^{+143}_{-588}$ & $55/61$\\
& NB &${2.1}_{-0.9}^{+6.9}$ & ${1.81}^{+0.12}_{-0.07}$ & ${1.5}^{+0.2}_{-0.5}$ & ${0.50}^{+0.01}_{-0.01}$ & ${0.11}^{+17.17}_{-0.01}$ & ${8.9}^{+0.3}_{-7.9}$ & ${1.06}^{+0.23}_{-0.01}$ & ${755}^{+32}_{-384}$ & $58/60$\\
& FB&${9.9}_{-8.4}^{+0.3}$ & ${1.40}^{+0.08}_{-0.04}$ & ${2.3}^{+2.3}_{-1.1}$ & ${7.6}^{+1.6}_{-7.0}$ & ${0.13}^{+0.72}_{-0.09}$ & ${12.1}^{+0.9}_{-5.4}$ & ${0.75}^{+0.15}_{-0.09}$ & ${3375}^{+3196}_{-2106}$ & $41/50$ \\
SL15 & HB &${7.8}_{-6.5}^{+1.7}$ & ${2.53}^{+0.03}_{-0.16}$ & ${4.2}^{+0.5}_{-0.5}$ & ${5.8}^{+3.3}_{-4.3}$ & ${10}^{+1}_{-8}$ & ${0.9}^{+2.7}_{-0.1}$ & ${1.04}^{+0.04}_{-0.12}$ & ${750}^{+585}_{-144}$ & $50/57$\\
& NB& ${1.8}_{-0.6}^{+6.8}$ & ${1.93}^{+0.08}_{-0.06}$ & ${2.4}^{+1.2}_{-0.8}$ & ${2.4}^{+6.8}_{-1.6}$ & ${0.68}^{+0.61}_{-0.44}$ & ${7.2}^{+1.6}_{-1.9}$ & ${1.08}^{+0.08}_{-0.11}$ & ${662}^{+418}_{-165}$ & $66/58$\\
& FB &${1.9}_{-0.5}^{+6.8}$ & ${1.45}^{+0.05}_{-0.06}$ & ${2.5}^{+0.8}_{-0.6}$ & ${9.6}^{+0.0}_{-9.1}$ & ${0.34}^{+2.41}_{-0.21}$ & ${10.5}^{+1.5}_{-6.1}$ & ${0.81}^{+0.12}_{-0.09}$ & ${2320}^{+2031}_{-1157}$ & $49/50$\\
SL16 & HB&${8.4}_{-6.6}^{+1.2}$ & ${2.22}^{+0.10}_{-0.06}$ & ${3.9}^{+0.7}_{-2.2}$ & ${10.0}^{+0.3}_{-9.1}$ & ${0.34}^{+1.06}_{-0.21}$ & ${9.3}^{+1.9}_{-4.2}$ & ${1.12}^{+0.10}_{-0.07}$ & ${586}^{+172}_{-202}$ & $93/66$ \\
& NB&${9.4}_{-6.7}^{+0.3}$ & ${2.28}^{+0.07}_{-0.11}$ & ${4.0}^{+0.7}_{-0.1}$ & ${4.5}^{+4.6}_{-1.0}$ & ${3.2}^{+3.4}_{-2.2}$ & ${3.0}^{+3.6}_{-1.4}$ & ${1.05}^{+0.06}_{-0.10}$ & ${812}^{+456}_{-179}$ & $94/59$ \\

SL17 & HB&${4.2}_{-2.9}^{+5.1}$ & ${2.29}^{+0.08}_{-0.07}$ & ${3.8}^{+0.3}_{-1.3}$ & ${2.6}^{+5.1}_{-2.1}$ & ${1.1}^{+4.5}_{-0.6}$ & ${5.7}^{+2.3}_{-3.5}$ & ${1.07}^{+0.26}_{-0.03}$ & ${715}^{+80}_{-406}$ & $64/65$ \\
& NB& ${2.1}_{-0.5}^{+7.3}$ & ${2.14}^{+0.16}_{-0.05}$ & ${4.0}^{+0.6}_{-0.2}$ & ${3.1}^{+1.7}_{-1.5}$ & ${16}^{+2}_{-10}$ & ${0.6}^{+2.2}_{-0.1}$ & ${0.97}^{+0.06}_{-0.08}$ & ${1179}^{+127}_{-286}$ & $78/65$\\
& FB&${9.4}_{-7.8}^{+0.11}$ & ${1.49}^{+0.04}_{-0.07}$ & ${2.5}^{+0.9}_{-1.2}$ & ${9.7}^{+0.4}_{-9.1}$ & ${0.3}^{+2.9}_{-0.2}$ & ${10.7}^{+1.9}_{-6.4}$ & ${0.86}^{+0.10}_{-0.13}$ & ${1750}^{+2232}_{-677}$ & $58/48$ \\
SL18 & HB &${2.3}_{-0.9}^{+7.0}$ & ${2.17}^{+0.01}_{-0.07}$ & ${3.5}^{+0.4}_{-1.6}$ & ${1.9}^{+6.9}_{-1.0}$ & ${0.54}^{+0.07}_{-0.42}$ & ${8.7}^{+2.8}_{-0.1}$ & ${1.07}^{+0.14}_{-0.02}$ & ${699}^{+42}_{-333}$ & $84/69$\\
& NB& ${6.2}_{-4.4}^{+3.3}$ & ${2.07}^{+0.08}_{-0.17}$ & ${4.0}^{+0.7}_{-1.1}$ & ${9.9}^{+0.3}_{-9.2}$ & ${0.6}^{+1.4}_{-0.3}$ & ${7.7}^{+16.3}_{-3.6}$ & ${1.04}^{+0.32}_{-0.09}$ & ${888}^{+454}_{-566}$ & $75/62$\\
SL19 & HB&${7.7}_{-6.4}^{+1.8}$ & ${2.42}^{+0.05}_{-0.11}$ & ${3.8}^{+0.8}_{-0.1}$ & ${3.5}^{+6.0}_{-0.6}$ & ${1.9}^{+0.7}_{-1.1}$ & ${3.5}^{+2.4}_{-0.7}$ & ${1.05}^{+0.04}_{-0.08}$ & ${644}^{+290}_{-97}$ & $64/65$ \\
& NB &${9.9}_{-8.1}^{+0.2}$ & ${2.06}^{+0.08}_{-0.12}$ & ${4.2}^{+0.5}_{-1.1}$ & ${4.7}^{+4.3}_{-3.5}$ & ${1.2}^{+1.0}_{-1.1}$ & ${5.0}^{+4.6}_{-1.6}$ & ${1.00}^{+0.27}_{-0.08}$ & ${981}^{+496}_{-620}$ & $59/56$\\
SL20 & HB&${1.10}_{-0.02}^{+6.10}$ & ${2.54}^{+0.02}_{-0.19}$ & ${3.7}^{+0.6}_{-0.2}$ & ${2.1}^{+6.7}_{-1.0}$ & ${3.4}^{+0.5}_{-2.7}$ & ${2.4}^{+3.5}_{-0.4}$ & ${0.93}^{+0.11}_{-0.02}$ & ${953}^{+135}_{-384}$ & $69/65$ \\
SL21 & HB& ${3.5}_{-2.3}^{+5.6}$ & ${2.09}^{+0.01}_{-0.17}$ & ${3.8}^{+0.8}_{-2.0}$ & ${2.1}^{+7.2}_{-1.4}$ & ${2.3}^{+0.2}_{-2.2}$ & ${11.5}^{+0.8}_{-3.0}$ & ${1.00}^{+0.19}_{-0.07}$ & ${1069}^{+459}_{-567}$ & $55/59$\\
& NB&${5.4}_{-4.0}^{+3.9}$ & ${2.05}^{+0.03}_{-0.13}$ & ${3.8}^{+0.9}_{-1.2}$ & ${1.7}^{+6.8}_{-1.0}$ & ${2.1}^{+0.7}_{-1.7}$ & ${3.4}^{+6.2}_{-0.3}$ & ${0.99}^{+0.14}_{-0.07}$ & ${1094}^{+469}_{-497}$ & $54/53$\\
SL22 & NB&${3.4}_{-2.1}^{+5.7}$ & ${1.59}^{+0.16}_{-0.04}$ & ${1.3}^{+2.1}_{-0.3}$ & ${7.3}^{+2.1}_{-6.7}$ & ${0.5}^{+17.3}_{-0.4}$ & ${7.8}^{+0.8}_{-7.1}$ & ${0.99}^{+0.23}_{-0.07}$ & ${972}^{+365}_{-523}$ & $54/55$ \\
& FB&${3.2}_{-1.8}^{+6.3}$ & ${1.31}^{+0.02}_{-0.11}$ & ${1.4}^{+2.6}_{-0.4}$ & ${1.0}^{+8.3}_{-0.3}$ & ${0.20}^{+0.23}_{-0.20}$ & ${12.5}^{+26.0}_{-1.0}$ & ${0.70}^{+0.15}_{-0.09}$ & ${4238}^{+4902}_{-1485}$ & $85/48$ \\
SL23 & NB&${1.02}_{-0.02}^{+8.00}$ & ${1.76}^{+0.18}_{-0.04}$ & ${1.3}^{+3.0}_{-0.3}$ & ${9.7}^{+0.5}_{-8.9}$ & ${0.7}^{+11.9}_{-0.6}$ & ${7.0}^{+0.5}_{-6.0}$ & ${1.10}^{+0.22}_{-0.03}$ & ${618}^{+81}_{-300}$ & $75/55$ \\
& FB &${8.1}_{-6.9}^{+1.4}$ & ${1.46}^{+0.06}_{-0.05}$ & ${3.5}^{+1.2}_{-2.2}$ & ${2.5}^{+6.6}_{-1.9}$ & ${0.12}^{+0.70}_{-0.10}$ & ${10.4}^{+1.8}_{-3.8}$ & ${0.81}^{+0.07}_{-0.18}$ & ${2236}^{+6752}_{-704}$ & $64/47$\\
SL24 & HB&${3.1}_{-1.8}^{+6.3}$ & ${2.33}^{+0.13}_{-0.17}$ & ${3.5}^{+1.2}_{-1.5}$ & ${0.8}^{+8.6}_{-0.2}$ & ${2.6}^{+0.1}_{-2.4}$ & ${3.7}^{+7.3}_{-0.5}$ & ${1.06}^{+0.19}_{-0.07}$ & ${787}^{+280}_{-419}$ & $71/60$ \\
& NB& ${1.3}_{-0.2}^{+6.0}$ & ${2.42}^{+0.09}_{-0.15}$ & ${3.7}^{+0.8}_{-0.2}$ & ${3.4}^{+6.1}_{-2.0}$ & ${8.3}^{+4.1}_{-6.5}$ & ${1.5}^{+3.0}_{-0.6}$ & ${0.97}^{+0.20}_{-0.06}$ & ${1304}^{+534}_{-729}$ & $75/60$\\
SL25 & HB&${3.4}_{-2.2}^{+6.0}$ & ${2.16}^{+0.24}_{-0.02}$ & ${1.9}^{+2.7}_{-0.3}$ & ${1.1}^{+8.1}_{-0.4}$ & ${0.3}^{+1.7}_{-0.1}$ & ${11.3}^{+0.4}_{-6.9}$ & ${1.10}^{+0.15}_{-0.10}$ & ${537}^{+604}_{-129}$ & $60/56$ \\
& NB&${9.8}^{+0.4}_{-8.2}$ & ${2.37}^{+0.99}_{-0.02}$ & ${3.9}^{+0.8}_{-0.2}$ & ${5.2}^{+3.7}_{-4.3}$ & ${3.9}^{+15.2}_{-0.7}$ & ${2.4}^{+0.3}_{-2.1}$ & ${1.10}^{+0.55}_{-0.05}$ & ${740}^{+175}_{-586}$ & $77/56$ \\

\bottomrule
\end{longtable}
\vspace{-5pt}
\noindent Note.-- The disk density $\log(n_e)$ is fixed at 19 cm$^{-3}$, spin is fixed at $a=0$, and emissivity index is fixed at $q=3$. The column density $N_H\ (10^{22}\ \rm cm^{-2})$ and inclination is fixed at the average values from the focused spectral analysis of ${6.39}$ and $34.08^\circ$, respectively. 

\begin{table*}[ht]
\begin{center}
\caption{Maximum BL radius for focused observations}
\label{tab:BL}
\scriptsize
\begin{tabular}{cccccc}
\toprule
obs\# & Branch  & \rin (\risco) &  \rin (\rg)

& Flux$_{\rm unabs}$ & $R_{max}\ (R_g)$\\
\midrule
F01 & LN & ${1.48}^{+0.07}_{-0.12}$ & ${8.9}^{+0.4}_{-0.6}$  &1.04 &11.0  \\
& SA & ${1.19}^{+0.11}_{-0.04}$ & ${7.1}^{+0.7}_{-0.2}$ &0.9 & 9.9 \\
& FB & ${1.48}^{+0.04}_{-0.23}$ & ${8.9}^{+0.2}_{-1.4}$& 1.06& 11.2\\
F02 & UN & ${1.17}^{+0.01}_{-0.05}$  & ${7.0}^{+0.1}_{-0.3}$&1.3 & 12.9 \\
& LN & ${1.4}^{+0.3}_{-0.2}$ & ${8.4}^{+1.8}_{-1.2}$ &1.08 & 11.3 \\
& SA & ${1.41}^{+0.19}_{-0.04}$ & ${8.5}^{+1.1}_{-0.2}$ &0.9 & 9.8 \\
& FB & ${1.34}^{+0.09}_{-0.14}$ & ${8.0}^{+0.5}_{-0.8}$ & 1.03&10.9\\
F03 & UN & ${1.12}^{+0.05}_{-0.03}$ & ${6.7}^{+0.3}_{-0.2}$ &1.32 &13.1 \\
& LN & ${1.22}^{+0.03}_{-0.17}$ & ${7.3}^{+0.2}_{-1.0}$ &1.09 & 10.4 \\
& SA & ${2.23}^{+0.08}_{-0.31}$ & ${13.4}^{+0.5}_{-1.9}$ & 0.9&10.3  \\
F04 & UH & ${1.49}^{+0.11}_{-0.08}$ & ${8.9}^{+0.7}_{-0.5}$ & 1.2&11.8 \\
F05 & UH & ${1.44}^{+0.08}_{-0.10}$ & ${8.6}^{+0.5}_{-0.6}$ &1.3 &12.9 \\
F06 & UH & ${1.48}^{+0.07}_{-0.15}$& ${8.9}^{+0.4}_{-0.9}$ &1.3 &13.3\\
& LH & ${1.3}\pm0.2$ & ${7.8}\pm1.2$ &1.6 &15.1 \\
F07 & UH & ${1.06}^{+0.03}_{-0.05}$ & ${6.4}^{+0.2}_{-0.3}$ &1.2 & 12.5 \\
& LH & ${1.02}^{+0.08}_{-0.02}$ & ${6.1}^{+0.5}_{-0.1}$ &1.6 & 15.8 \\
\bottomrule
\hline
\end{tabular}
\end{center}
Note. -- The values for \rin are obtained from reflection modeling in Table~\ref{tab:reflection_focused}. Values in \risco are converted to \rg given that the spectral fits were conducted with the dimensionless spin parameter set as $a=0$ which corresponds to 1 \risco = 6 \rg. The mass accretion rate is calculated from the unabsorbed flux (Flux$_{\rm unabs}$) from 0.5--50 keV (given in units of $10^{-8}$ \fluxcgs) assuming a distance of 11 kpc and an accretion efficiency of $\eta =0.2$. The canonical values are used for NS mass and radius ($1.4\ M_{\odot}$, 10 km).
\end{table*}

\FloatBarrier
\begin{figure*}[ht]
    \centering
    \includegraphics[width=\textwidth]{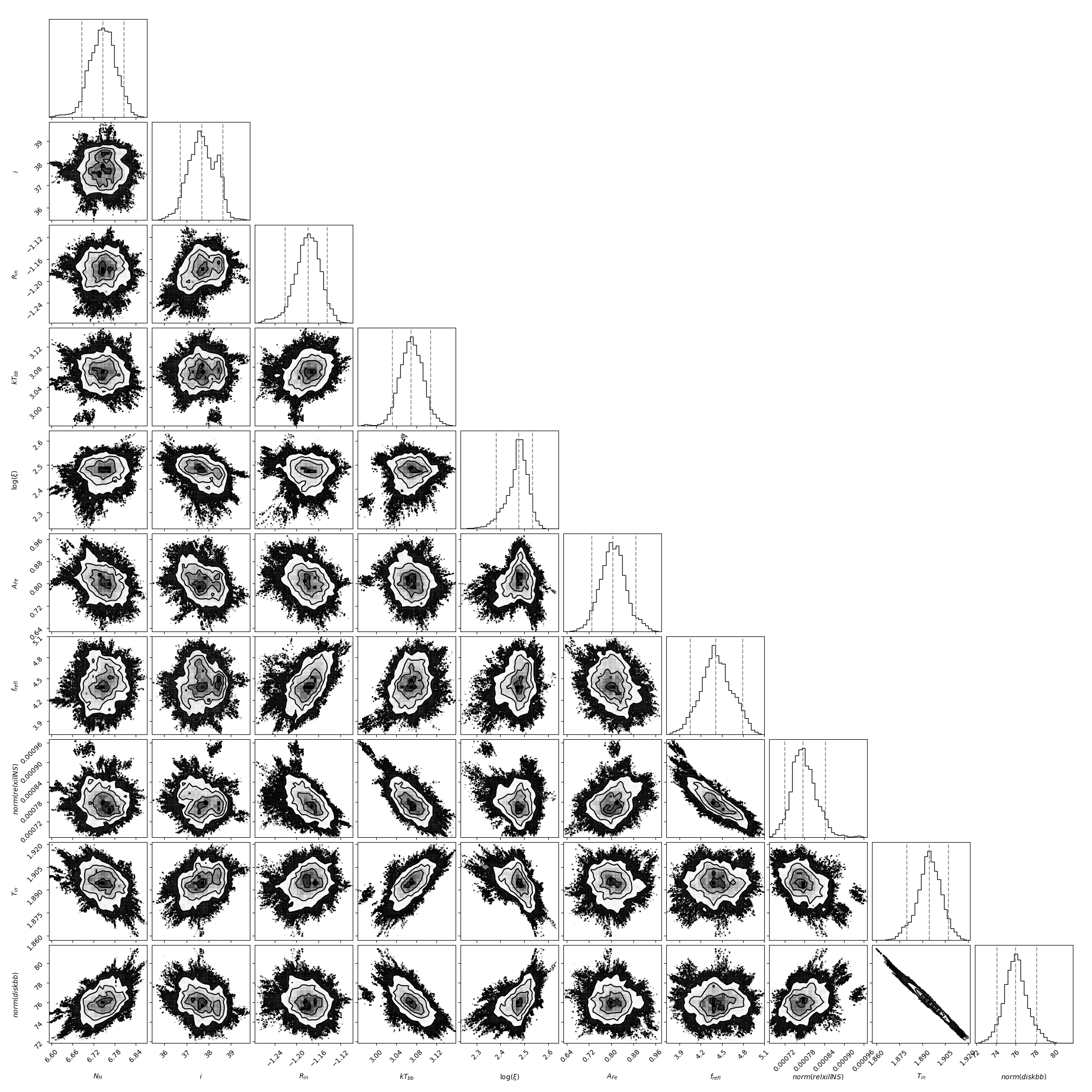}
    \caption{Corner plot for F02A\_UN. Grey dashed line is the 90\% confident level. The disk ionization parameter is not degenerate with the reflection fraction, thus the correlation we see between ionization and reflection fraction is likely due to a physical correlation between the two parameters in the system.}
    \label{fig:corner}
\end{figure*}



\end{document}